\documentclass[11pt,a4paper]{article}
\pdfoutput=1
\usepackage{jheppub}

\usepackage{color}
\usepackage{amsmath}
\usepackage{pifont}   
\usepackage{verbatim}   
\usepackage{acronym} 
    
\usepackage{amsfonts}
\usepackage{amssymb}
\usepackage{mathrsfs} 
\usepackage{graphicx}
\usepackage{multirow} 
\usepackage{slashed} 
\usepackage{array}

\usepackage{graphicx}
\usepackage{epsfig}
\usepackage{lscape}
\usepackage{rotating}
\usepackage{epsf}
\usepackage{bm}
\usepackage{booktabs}
\usepackage{array}
\usepackage{caption}
\usepackage{subcaption}
\usepackage[utf8]{inputenc}
\usepackage{float}

\usepackage{multirow}
\usepackage{array,booktabs,colortbl,multirow}
\usepackage{colortbl,xcolor,colordvi,color}
\usepackage{rotating}
\allowdisplaybreaks

\newcommand{\beq}{\begin{align}}% can be used as {equation} or  {align}
\newcommand{\eeq}{\end{align}}
\def\be{\begin{equation}}
\def\ee{\end{equation}}
\def\bea{\begin{eqnarray}}
\def\eea{\end{eqnarray}}
\def\bitem{\begin{itemize}}
\def\eitem{\end{itemize}}
\newcommand{\bec}{\begin{center}}
\newcommand{\eec}{\end{center}}
\newcommand{\ba}{\begin{array}}
\newcommand{\ea}{\end{array}}
\newcommand{\cmrule}{\midrule[0.25mm]}

\usepackage{bbm}

\title{Collider phenomenology of Hidden Valley mediators of spin 0 or 1/2 with semivisible jets}

\affiliation{Instituto de F\'isica, Universidade de S\~ao Paulo, \\C.P. 66.318, 05315-970 S\~ao Paulo, Brazil}

\author{Hugues Beauchesne,}
\author{Enrico Bertuzzo,}
\author{Giovanni Grilli di Cortona, and}
\author{Zahra Tabrizi}

\emailAdd{hubea44@if.usp.br, bertuzzo@if.usp.br, ggrilli@if.usp.br, ztabrizi@if.usp.br}

\abstract{Many models of Beyond the Standard Model physics contain particles that are charged under both Standard Model and Hidden Valley gauge groups, yet very little effort has been put into establishing their experimental signatures. We provide a general overview of the collider phenomenology of spin 0 or 1/2 mediators with non-trivial gauge numbers under both the Standard Model and a single new confining group. Due to the possibility of many unconventional signatures, the focus is on direct production with semivisible jets. For the mediators to be able to decay, a global $U(1)$ symmetry must be broken. This is best done by introducing a set of operators explicitly violating this symmetry. We find that there is only a finite number of such renormalizable operators and that the phenomenology can be classified into five distinct categories. We show that large regions of the parameter space are already excluded, while others are unconstrained by current search strategies. We also discuss how searches could be modified to better probe these unconstrained regions by exploiting special properties of semivisible jets.}

\begin{document}

\maketitle
\section{Introduction}\label{Sec:Intro}
As the LHC continues its operation, the most sought after models of Beyond the Standard Model (BSM) physics will have to hide in an increasingly small window if they are to remain both unexcluded and potentially discoverable at that collider. It is therefore relevant to look for signatures of BSM physics that might have evaded current constraints simply because we did not search for them or because they did not fit into our preconceived notions of what BSM physics should look like.

A great source of such signatures is Hidden Valley models \cite{Strassler:2006im, Strassler:2006ri, Harnik:2008ax}. This ubiquitous class of models is characterized by a new low energy confining sector that communicates with the Standard Model (SM) via some heavy mediators. Up to now, most studies of the collider phenomenology of Hidden Valley models have focused on spin 1 mediators (e.g. \cite{Han:2007ae, Strassler:2008fv, Baumgart:2009tn, Seth:2011ci, Chan:2011aa, Strassler:2006im, Pierce:2017taw}). However, Hidden Valley mediators of spin 0 or 1/2 also appear in a multitude of models. For example, UV completions of Twin Higgs models \cite{Chacko:2005pe} can include fermions that are charged under both Standard Model electroweak groups and mirror QCD. They even represent one of the best directions to discover these models \cite{Cheng:2016uqk}. In Folded SUSY \cite{Burdman:2006tz}, there can be top partners with similar gauge numbers but with spin 0 instead. Additionally, fermions charged under both electroweak gauge groups and a new strong group have been used in Relaxion models to generate the back reaction potential \cite{Graham:2015cka}. Perhaps more importantly, mediators of spin 0 or 1/2 appear in several models of composite dark matter or their UV completions (see~\cite{Cline:2013zca, Lewis:2014boa, Antipin:2015xia, Kribs:2016cew} for several reviews).

If these mediators are ever directly produced and assuming they are unstable, they will generally decay to a particle that is a singlet under all Standard Model groups but still charged under a new strong group. The resulting dark hadronization will lead to the production of dark hadrons. Typically, some will be unstable and some will be stable because of some symmetry of the dark sector. The decay length of the unstable dark hadrons can vary wildly, which leads to completely different collider phenomenology. If they are stable on collider scales, the dark hadrons will simply leave the detector and be counted as Missing Transverse Energy (MET). The signature is then not so different from Supersymmetry (SUSY) with $R$-parity. If instead the unstable dark hadrons decay inside the detector, the collider signature depends on the dark hadron multiplicity. Low multiplicity leads to displaced vertices, for which there are already several experimental searches (e.g. \cite{King:2016ort, Aad:2015uaa}). The constraints in this case are mainly on the cross section and decay length, with the details of the mediator affecting only subdominant effects like the boost and pseudorapidity of the dark hadrons. Higher multiplicity leads to emerging jets \cite{Schwaller:2015gea}, which are also mostly independent of the details of the mediators and not currently well constrained by collider searches. Finally and arguably more interestingly, promptly decaying dark hadrons lead to semivisible jets \cite{Cohen:2015toa}. These objects appear like normal jets, except that they are accompanied by some collinear MET originating from the stable dark hadrons. This MET is easy to overlook, as it can readily be mistaken for an effect of the inevitably imperfect energy resolution. This is specially crucial for search strategies based on only jets and MET. Previous searches on the collider phenomenology of semivisible jets include Refs.~\cite{Cohen:2015toa, Cohen:2017pzm}.

With these considerations in mind, the purpose of the present article is to investigate direct production of Hidden Valley mediators of spin 0 or 1/2 that are non-trivially charged under both the Standard Model gauge groups and a single new confining group. Because of their unique signatures, we focus on the case where the mediators decay to semivisible jets. This article is meant to serve as a guide that is reasonably independent of model assumptions, but which can be applied to a wide class of models. We follow the spirit of previous similar guides like Refs.~\cite{Diaz:2017lit, Gunion:1989we, DeSimone:2012fs}. More precisely, we seek to provide an overview of current experimental constraints on spin 0 or 1/2 mediators and show that some regions of parameter space are still relatively unconstrained. We also discuss how current search strategies could be modified to account for these unconstrained regions. This will mostly rely on the ability to infer the expected direction of the MET from other reconstructed objects. A special emphasis is put on the average fraction of the energy of a given semivisible jet that is transmitted to MET, which we refer to as $r_{\text{inv}}$. A side effect of studying this parameter is that taking the limit $r_{\text{inv}}=1$ leads to constraints that can be applied to mediators stable on collider scales.

For generic gauge numbers of the mediator, the Lagrangian respects an accidental $U(1)$ symmetry which renders the mediator stable and which must be broken to lead to conventional Hidden Valley phenomenology. Even though this symmetry could be broken spontaneously, it would lead to several potential problems and we instead introduce operators that break it explicitly. We find that there is only a finite number of such renormalizable operators and that the phenomenology can consequently be classified into five distinct categories. In case I, the mediator decays to a semivisible jet and a normal jet. For small $r_{\text{inv}}$, paired dijets searches provide decent coverage, while large $r_{\text{inv}}$ is well constrained by searches for squarks and gluinos with large MET. There is however a region of intermediary $r_{\text{inv}}$ relatively less constrained. In case II, the mediator decays to a semivisible jet and a lepton. The case of large $r_{\text{inv}}$ is well constrained by slepton searches, but most bounds vanish when $r_{\text{inv}}$ drops. Cuts on the direction of the MET could however be very efficient at distinguishing the signal from the background for intermediary values of $r_{\text{inv}}$. In case III, a fermion mediator and a Hidden Valley fermion interact with the Higgs, which leads to mixing between these fermions and decays involving massive gauge bosons or the Higgs. The case of large $r_{\text{inv}}$ can be well constrained by SUSY electroweakino searches, while low $r_{\text{inv}}$ is only poorly constrained by measurements of the $WZ$ differential cross section. Cuts on the direction of the MET can also be applied, but with lesser success than in case II. Case IV consists of operators linear in the mediator and that contain only scalars. These are mainly constrained by Higgs physics and Electroweak Precision Test (EWPT). Case V represents operators that depend non-linearly on the mediator. They lead to a great variety of exotic signatures.

This article is organized as follow. We first discuss a few simplifying assumptions and list all possible operators. Our simulation procedure is then explained. We next proceed to discuss the operators of category I, II, III, IV and V in that order. We reserve the last section for concluding remarks. The article is structured such that each section presenting a category can be read independently of the others.

\section{Overview of assumptions, operators and simulation procedure}\label{Sec:AssumptionListOperators}
The large number of possible Hidden Valley models renders a complete description of a general dark sector practically unfeasible. Fortunately, many of the details of such a sector are mostly irrelevant when it comes to collider constraints. Because of this and for the sake of alleviating the presentation, we begin by listing a few reasonable assumptions we will make throughout the article. Our work will mimic in a lot of ways supersymmetric simplified models \cite{Alves:2011wf}. The present section also includes the necessary conditions for the mediator to actually be able to decay to the Hidden sector. This will require the introduction of operators of which we present the complete list. We also discuss our implementation of collider simulation.

\subsection{General setup}\label{sSec:GeneralAssumptions}
We limit ourselves to studying cases where only one mediator field is present at a time. When the mediator is charged under a broken symmetry like $SU(2)_L$, this will result in different particles which we will all consider. We assume the presence of a single new confining group $\mathcal{G}$. Its confinement scale $\Lambda$ is assumed low enough to be probed at the LHC, yet above the QCD scale. Some of the operators presented below will lead to $\mathcal{G}$ being spontaneously broken, but we assume no other source of breaking exists otherwise.

For conceptual purposes, we will refer to the mediator in general as $X$. In the different cases presented below, we will add additional subscripts and superscripts to indicate their exact gauge numbers or particles they couple to. Under the basic assumptions of this article, we assume $X$ to be non-trivially charged under both the Standard Model gauge groups and $\mathcal{G}$. It can both be a scalar (superscript $S$) or a fermion (superscript $F$). Fermion mediators are assumed to be vector fermions. Scalars will be forced to be complex with a few possible exceptions. To ascribe to the commonly understood notion of Hidden Valley models, we assume the mass of $X$ to be much larger than the confinement scale of $\mathcal{G}$. When $\mathcal{G}$ is broken, we will refer to the mediators charged under the unbroken subgroup as $X^U$ and the others as $X^B$.

We also introduce a particle charged only under $\mathcal{G}$, which we refer to in general as $n$. It represents a dark (s)quark and will appear as a decay product of $X$. Its scalar iteration will be referred to as $n^S$ and its fermion iteration as $n^F$. In the fermion case, we take it to be a vector fermion. Scalars will in general be complex with a few possible exceptions. We assume $n$ to be much lighter than the mediator $X$, again to ascribe to the common Hidden Valley paradigm. We also assume $\Lambda$ not to be too low compared to the mass of $n$, as this would lead to Hidden sector quirks \cite{Kang:2008ea} and is beyond the scope of the article. There could in principle be many generations of $n_i$ and this could be used to accommodate a large number of residual symmetries, which could explain a large $r_{\text{inv}}$. For example, assume a large number of generations of $n_i$ and that $X$ only communicates with one of them. Unless there is some symmetry breaking in the dark sector, there is a set of $U(1)$ symmetries associated to each dark generation. Dark mesons such as $\bar{n}_i n_j$ would be unstable for $i=j$ and stable for $i\neq j$. Assuming all generations to be similar in mass, one would then expect the production of a large number of stable mesons and therefore a large $r_{\text{inv}}$. For our analysis, we will concentrate on the case of a single $n$, while considering $r_{\text{inv}}$ a free parameter and keeping in mind that multiple generations would in principle be possible. Under our assumptions, semivisible jets will originate from $X$ decay and will therefore be very boosted. To a good approximation, the only important factor is then $r_{\text{inv}}$ and beyond this the number of generations of $n_i$ is mostly irrelevant. When $\mathcal{G}$ is broken, dark (s)quark that are charged under the unbroken subgroup will be referred to as $n^U$ and those that are not as $n^B$.

It is worth mentioning that in principle the fraction of the energy of a semivisible jet that is transmitted to MET could be dependent on its energy. Even at the same energy, there should also be statistical fluctuations that are dependent on the exact details of the dark sector. The approach we will take is to ignore the energy dependence, but to study variance of the energy transmitted to MET when relevant. The parameter $r_{\text{inv}}$ is then more properly defined as the energy independent average fraction of the energy of a semivisible jet that is converted to MET.

The last particles to be introduced are the gauge bosons of $\mathcal{G}$, which we label $g_D$. When $\mathcal{G}$ is broken, we will refer to the gauge bosons which are charged under an unbroken subgroups as $g_D^U$ and those that are not as $g_D^B$.

\subsection{Decay of the mediator}\label{sSec:DecayMediator}
To lead to conventional Hidden Valley phenomenology, a given mediator must satisfy two conditions: It should in principle be possible to produce it at colliders and it should be able to decay to $n$.\footnote{We refer to Ref.~\cite{Li:2017xyf} for the phenomenology of stable mediators.} The first condition is trivially met, as we assume $X$ to be charged under Standard Model gauge groups. The second condition is however less trivial to satisfy. This has to do with the fact that, for general gauge numbers, $X$ can only appear in the Lagrangian multiplied by its conjugate. Such terms respect a global $U(1)$ symmetry, which we label $U(1)_X$ and which renders the mediators stable. It is then necessary to either break the symmetry via the introduction of an $U(1)_X$ violating operator or break it spontaneously. The latter is unfortunately rather problematic. The reason is that it would lead to an uneaten Goldstone boson that interacts with the gauge bosons of the Standard Model. In addition to obvious cosmological problems, it would lead to very dangerous decays and would be a major challenge for EWPT. One could of course assume $U(1)_X$ to be gauged, but the resulting massive gauge boson would still be light and cause considerable issues with EWPT. In addition, the possibilities for the mediator would be very limited, as it would need to be a scalar and be able to acquire a vacuum expectation value (vev) in such a way as to preserve both QCD and the electromagnetic group. We do emphasize that spontaneous breaking of $U(1)_X$ is not necessarily entirely ruled out, but simply more difficult and less attractive.

Operators that break $U(1)_X$ are however a perfectly valid option and will be the focus of the article. They can contain multiple insertions of $X$ or its conjugate. Operators involving only one $X$ obviously break the $U(1)_X$ symmetry. Under our assumptions, they are forced to also contain at least one SM field and at least one $n$. When fermions are involved, dimensional analysis forces $X$ to have the same SM gauge number as another SM field. It is then easy to list all the possible gauge numbers. When the operator only involves scalars, it is a simple combinatoric exercise to list all the ways we can combine one $X$ and at least one $H$ and one $n$. Those that contain twice the mediator $X$ unavoidably respect a residual $\mathbb{Z}_2$ symmetry that leaves the mediators stable. These operators do not serve their required purpose and we will ignore them from now on. The same comment applies to operators involving four iterations of $X$. Finally, operators can contain three copies of $X$ or its conjugate. Those containing $XXX$ or $X^\dagger X^\dagger X^\dagger$ respect a discrete $\mathbb{Z}_3$ subgroup that leaves the mediators stable. We will once again ignore these operators. The $U(1)_X$ symmetry can however be broken if the operator contains twice the mediator and once its conjugate, i.e. $XXX^\dagger$. These operators break $U(1)_X$ and do not allow for any unbroken discrete symmetry. Non-renormalizable operators will not be considered.\footnote{Note that it would be possible to introduce operators that break $U(1)_X$ to one of its discrete subgroup and combine this with spontaneous breaking of $U(1)_X$. Note also that a stable mediator could still lead to semivisible jets, albeit at loop level. Though interesting, these points are beyond the scope of this article.}

Note that real mediators would instead respect a discrete $\mathbb{Z}_2$ symmetry. This symmetry would then need to be broken by operators that are either linear in $X$ or proportional to $X^3$. These operators will turn out to be simple subcases of the operator that break $U(1)_X$, except of course that the mediator is now assumed to be real.

Under these considerations, the operators can be classified into five distinct categories which we now present. In a similar fashion to $X$ and $n$, the coefficients of these operators will in general be referred to as $\lambda$, with additional subscripts and superscripts when needed.

\subsection*{Case I: Three field operators involving quarks}
The first category of operators is those that include a quark field. They are:
\begin{equation}\label{eq:OperatorsC1}
  \begin{aligned}
    &\lambda^F_{Q_i}   (n^S)^\dagger \bar{X}^F_{Q}  P_L Q_i    +\text{h.c.} & \;\;\;\;\;
    &\lambda^F_{U^c_i} (n^S)^\dagger \bar{X}^F_{U^c}P_R U^c_i  +\text{h.c.} & \;\;\;\;\;
    &\lambda^F_{D^c_i} (n^S)^\dagger \bar{X}^F_{D^c}P_R D^c_i  +\text{h.c.}\\
    &\lambda^S_{Q_i}   (X^S_{Q})  ^\dagger \bar{n}^F P_L Q_i   +\text{h.c.} &
    &\lambda^S_{U^c_i} (X^S_{U^c})^\dagger \bar{n}^F P_R U^c_i +\text{h.c.} &
    &\lambda^S_{D^c_i} (X^S_{D^c})^\dagger \bar{n}^F P_R D^c_i +\text{h.c.} &
  \end{aligned}
\end{equation}
The fields $Q$, $U^c$ and $D^c$ are the usual SM quark fields written as four components spinors. In this notation, $X$ has the same SM gauge numbers as the corresponding quark. We will generally assume $X$ to be a fundamental of $\mathcal{G}$ for simplicity, but nothing would in principle prohibit higher representations. The subscripts on $X$ and $\lambda$ correspond to the SM particles $X$ interacts with. The superscript of $\lambda$ corresponds to the spin of $X$. The particle $n$ is required to be an antifundamental of $\mathcal{G}$. There is an implicit summation over the indices $i$. In practice, several non-zero indices of $\lambda$ would be dangerous for flavor physics and we simply assume that a single one is non-zero in the mass eigenstate basis. These operators do not lead to baryon number violation by themselves as the mediators can be assigned coherently a baryon number. Similar operators appeared in Ref.~\cite{Carloni:2010tw}, but the full left-right structure was not studied completely. This comment also applies to case II.

\subsection*{Case II: Three field operators involving leptons}
The second category of operators is similar to case I, but with quarks replaced with leptons. They are:
\begin{equation}\label{eq:OperatorsC2}
  \begin{aligned}
    &\lambda^F_{L_i}   (n^S)^\dagger \bar{X}^F_{L}  P_L L_i    +\text{h.c.} & \;\;\;\;\;
    &\lambda^F_{E^c_i} (n^S)^\dagger \bar{X}^F_{E^c}P_R E^c_i  +\text{h.c.} \\
    &\lambda^S_{L_i}   (X^S_{L})^\dagger   \bar{n}^F P_L L_i   +\text{h.c.} &
    &\lambda^S_{E^c_i} (X^S_{E^c})^\dagger \bar{n}^F P_R E^c_i +\text{h.c.} 
  \end{aligned}
\end{equation}
The fields $L$ and $E^c$ are the usual SM lepton fields. The rest of the notation is analogous to case I. We again assume the mediator communicates with only a single generation to avoid flavor issues. A lepton number can also be assigned coherently to these mediators.

\subsection*{Case III: Operators involving fermion mediators and the Higgs}
The third category is similar to the previous two, but with fermion $n$ and $X$ now interacting with the Higgs through a three field operator. Similar scalar mediators could also be included in this category, but their phenomenology is more closely related to the other operators of case IV. Case III is special in that two operators are allowed for the same gauge numbers of the mediator. The relevant Lagrangian is then:
\begin{equation}\label{eq:OperatorC3F}
  \mathcal{L} = \lambda_{H,L}^F H^\dagger \bar{n}^F P_L X^F_H + \lambda_{H,R}^F H^\dagger \bar{n}^F P_R X^F_H + \text{h.c.}
\end{equation}
The mediator $X_H^F$ is assumed to be an $SU(2)_L$ doublet of weak hypercharge $+1/2$. The fields $X^F_H$ and $n^F$ are both assumed fundamentals of $\mathcal{G}$, but other representations are also possible.

\subsection*{Case IV: Operators linear in $X$ involving only scalars}
The fourth category of operators are those linear in $X$ involving only scalars. There are four different possibilities. First, there is the operator:
\begin{equation}\label{eq:OperatorC4s1}
  \lambda_{4s1}  H^\dagger (n^S)^\dagger X^S_{4s1} + \text{h.c.}
\end{equation}
In this case, $X^S_{4s1}$ is a doublet of $SU(2)_L$ with weak hypercharge of $+1/2$. Both $X^S_{4s1}$ and $n^S$ are assumed fundamental of $\mathcal{G}$, but other representations are possible. Note that this is one of only two cases where $\lambda$ has units of energy. The second operator is:
\begin{equation}\label{eq:Operator4s2}
  \lambda_{4s2} H^\dagger (n^S)^\dagger X^S_{4s2} n^S +\text{h.c.}
\end{equation}
The mediator $X^S_{4s2}$ is also a doublet of $SU(2)_L$ with weak hypercharge of $+1/2$. The fields $X^S_{4s2}$ and $n^S$ must however transform differently under $\mathcal{G}$, though there are many possibilities. The simplest is that $n^S$ is a fundamental and $X^S_{4s2}$ an adjoint, which we will assume from now on. The third operator is:
\begin{equation}\label{eq:Operator4s3}
  \lambda_{4s3} (n^S)^\dagger H^\dagger X^S_{4s3} H +\text{h.c.},
\end{equation}
where
\begin{equation}\label{eq:tripletX1}
  X^S_{4s3} = (X^S_{4s3})^\alpha \sigma^\alpha/2.
\end{equation}
The mediator $X^S_{4s3}$ is then a triplet of $SU(2)_L$ with no hypercharge. Both $X^S_{4s3}$ and $n^S$ are assumed fundamentals of $\mathcal{G}$, though other representations are possible. Note that $X^S_{4s3}$ and $n^S$ can be taken real if they transform under a real representations of $\mathcal{G}$. Finally, the last operator is:
\begin{equation}\label{eq:Operator4s4}
  \lambda_{4s4} (n^S)^\dagger H^T X^S_{4s4} H +\text{h.c.},
\end{equation}
with again
\begin{equation}\label{eq:tripletX3}
  X^S_{4s4} = (X^S_{4s4})^\alpha \sigma^\alpha/2.
\end{equation}
The mediator $X^S_{4s4}$ is again a triplet of $SU(2)_L$, but now with a weak hypercharge of $-1$. Both $X^S_{4s4}$ and $n^S$ are still assumed fundamentals of $\mathcal{G}$.

Note that it is possible to write a few additional operators by using $\tilde{H}=i\sigma_2 H^*$, but such operators can easily be proven to be equivalent to the ones written above by field redefinition. If $\mathcal{G}$ corresponded to a dark $SU(2)$, it would be possible to include additional operators. Their SM numbers would however be the same and the collider phenomenology would be unchanged under our assumptions. Note also that there exists an operator of the form $\lambda H^\dagger n^S X n^S +\text{h.c.}$ As far as collider phenomenology is concerned, it is however equivalent to the one of Eq.~(\ref{eq:Operator4s2}).

\subsection*{Case V: Operators non-linear in $X$}
Finally, there are three operators that are non-linear in $X$. Due to their nature, the mediators will be able to take on a multitude of gauge numbers. We will discuss this more carefully in Sec.~\ref{Sec:CaseV}. The operators are:
\begin{equation}\label{eq:Operatornl1}
  \lambda_{nl1}X^S_{nl1}X^S_{nl1}(X^S_{nl1})^\dagger H + \text{h.c.}
\end{equation}
\begin{equation}\label{eq:Operatornl2}
  \lambda_{nl2}X^S_{nl2}X^S_{nl2}(X^S_{nl2})^\dagger n + \text{h.c.}
\end{equation}
\begin{equation}\label{eq:Operatornl3}
  \lambda_{nl3}X^S_{nl3}X^S_{nl3}(X^S_{nl3})^\dagger + \text{h.c.}
\end{equation}
Note that the some of these operators allow for $X$ and $n$ to be real scalar fields. In this case, the dagger and the Hermitian conjugate can be omitted. Note also that $\lambda_{nl3}$ has units of energy.

Lastly, we note that there would be many additional possibilities if $X$ were not charged under $\mathcal{G}$. This includes scalar diquarks, dileptons, leptoquarks and a heavier Higgs doublet. Although interesting, they are outside the scope of this article. Besides these exceptions, the list of spin 0 or 1/2 mediators is complete to the best of our knowledge.

\subsection{Simulation procedure}\label{sSec:SimulationProcedure}
In the upcoming sections, we will presents bounds that were obtained by collider simulations. Signals are generated using \texttt{MadGraph5\_aMC@NLO} \cite{Alwall:2014hca} and an implementation of the different cases in Feynrules \cite{Alloul:2013bka}. The dark hadronization is simulated by \texttt{PYTHIA}8 \cite{Sjostrand:2007gs} using the Hidden Valley scenario \cite{Carloni:2011kk, Carloni:2010tw}. The patch of Ref.~\cite{Schwaller:2015gea} was implemented to properly take into account running of the gauge coupling constant of $\mathcal{G}$. Unless stated otherwise, $\mathcal{G}$ is taken to be $SU(3)$, the dark confinement scale is set to 10~GeV and only one fundamental $n$ is assumed to contribute to the running. A dark confinement scale close to the QCD scale is not an artificial choice as it can arise in many BSM models (e.g Twin Higgs). It is however impossible to completely decouple all effects of the confinement scale and we will discuss the repercussions of changing its value when relevant. Following Ref.~\cite{Cohen:2015toa}, semivisible jets are simulated by decaying a fraction $r_{\text{inv}}$ of all dark mesons to invisible particles and the rest to hadronic states. The mass of the lightest dark meson is set equal to 10~GeV. The effects of varying this mass are studied in appendix~\ref{app:variations}. \texttt{Delphes} 3 \cite{deFavereau:2013fsa} is used to simulate detector effects. Jets are reconstructed using \texttt{FastJet} \cite{Cacciari:2011ma} and the anti-$k_T$ algorithm \cite{Cacciari:2008gp}. Cross sections are calculated with \texttt{MadGraph5\_aMC@NLO} including next-to-leading order QCD corrections. 

\section{Case I: Three field operators involving quarks}\label{Sec:CaseI}
Case I consists of three field operators involving a quark, the mediator and a dark (s)quark. At hadron colliders, the mediators $X$ are typically pair produced via strong interactions, resulting in large cross sections. Diagrams depending on $\lambda$ also contribute to the cross section, but we take this parameter to be small to ease the presentation and we thus neglect them. This also suppresses the cross section for the pair production of $n$ and for production of $X$ in conjunction with $n$. We choose to focus on pair production of $X$ because it is a process that will always be important no matter the value of $\lambda$. Once produced, the mediators will each decay promptly to a semivisible jet and either a light jet, a b-jet or a top quark. Depending on $r_{\text{inv}}$, there are several possible signatures: large $r_{\text{inv}}$ leads to two jets and MET, intermediate $r_{\text{inv}}$ leads to four jets and MET (namely two jets and two semivisible jets), while small $r_{\text{inv}}$ gives rise to signatures with only four jets. 

\begin{figure}[t]
\begin{center}
    \includegraphics[width=0.6\textwidth, viewport = 0 0 450 400]{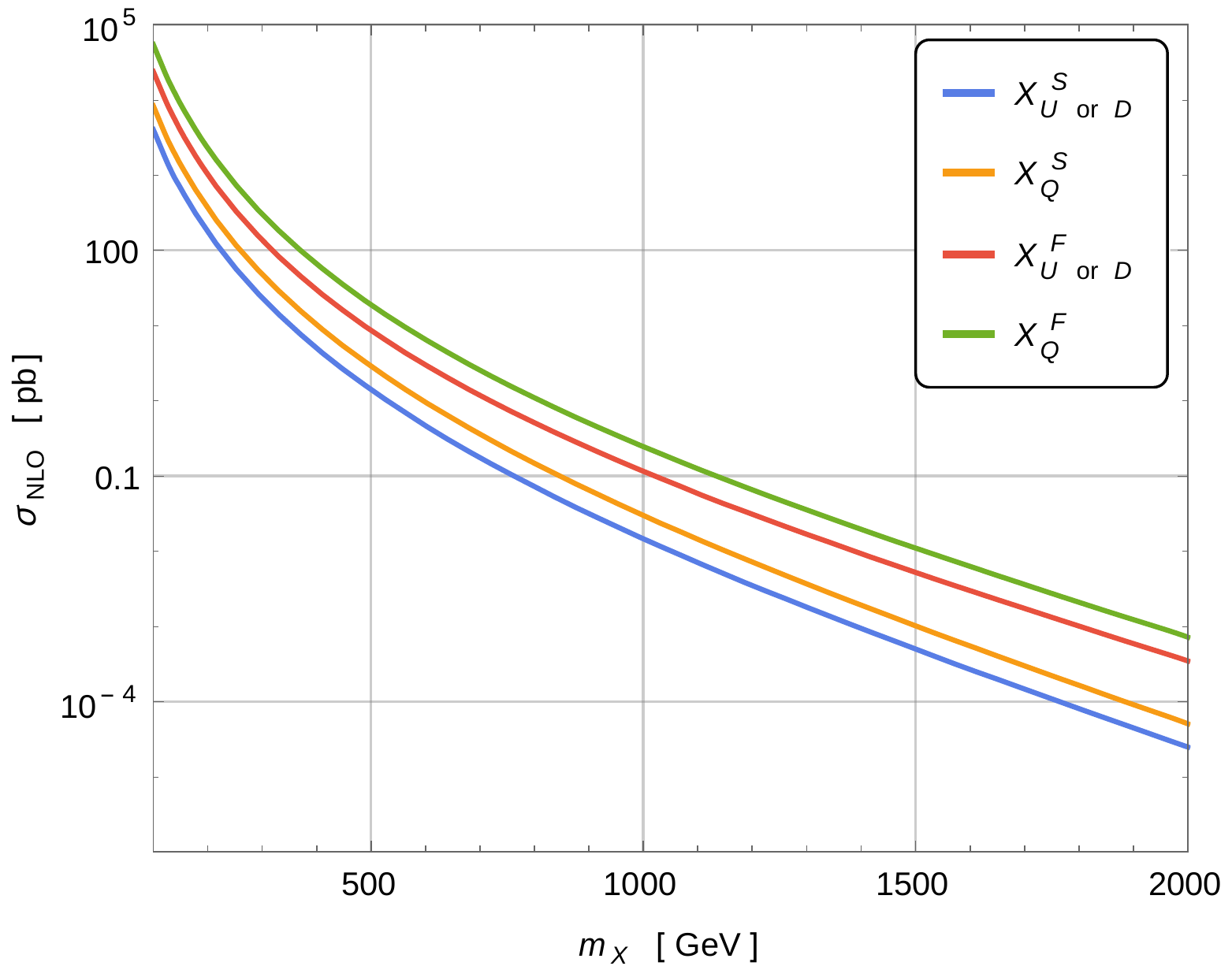}
  \end{center} 
\caption{NLO cross sections for mediators of category I. The Hidden Valley group $\mathcal{G}$ is taken to be $SU(3)$.} \label{fig:CaseI_CS_NLO}
\end{figure}

Fig.~\ref{fig:CaseI_CS_NLO} shows the cross section for pair production of mediators of category I as a function of their mass for $\mathcal{G}$ corresponding to $SU(3)$. It is worth noting that the cross sections for pair production of $X_{D^c}^S$ and $X_{U^c}^S$ are similar to those of squarks,\footnote{Note that many experimental searches assume mass degeneracy for the squarks of the first few generations. This results in a larger cross section and explains why some of our bounds will appear weaker than those of ATLAS or CMS.} while $X_{Q}^F$ has a cross section comparable to that of gluinos. We would therefore expect stronger constraints on the mass of this mediator.

In the next subsections, we will estimate and discuss current experimental constraints on mediators of category I for large and small $r_\text{inv}$.

\subsection{Case I with large $r_{\text{inv}}$}\label{sSec:CaseILargerinv}
If $r_{\text{inv}}$ is large, the experimental signatures of case I mediators can be classified into two different categories:
\begin{enumerate}
\item 2 high $p_T$ jets/b-jets/tops and MET
\item 2 high $p_T$ jets/b-jets/tops, 2 low $p_T$ jets and MET
\end{enumerate}
The larger $r_{\text{inv}}$, the larger is the fraction of dark hadrons that are stable for collider purposes and contribute to MET. In the limit of $r_\text{inv}=1$, all dark hadrons produced escape the detector. This leads to  2 high $p_T$ jets and MET. This is equivalent to squark pair production in Supersymmetry with $R$-parity conserved. As $r_{\text{inv}}$ decreases, the fraction of stable dark hadrons becomes smaller, leading to 2 high $p_T$ jets, 2 low $p_T$ jets and some MET. This is similar to gluino pair production, with the gluinos decaying to two jets and a stable neutralino (in a model with an effective coupling involving the gluino, the neutralino and two quarks).

We obtain collider bounds for mediators coupling to the first two generations and large $r_{\text{inv}}$ by recasting the ATLAS search of Ref.~\cite{ATLAS:2017cjl}. The data was obtained at $\sqrt{s}=13$ TeV with 36.1~fb$^{-1}$ of integrated luminosity. The experimental search in Ref.~\cite{ATLAS:2017cjl} presents two different approaches: the first one uses an effective mass based search, where $M_\text{eff}$ is defined as the scalar sum of the transverse momenta of all the jets with $p_T> 50 $ GeV; the second one uses Recursive Jigsaw Reconstruction techniques \cite{Buckley:2013kua,Jackson:2016mfb}, specific for models with compressed spectra. We concentrate on the signal regions of the $M_\text{eff}$ based search, since they give the strongest limits for small neutralino masses. 

More specifically, the signal regions consist of two sets of regions that target different classes of signals. Eight signal regions aim at squark pair production, requiring at least two or three jets with different cuts on $M_\text{eff}$ and MET. Different thresholds on $M_{\text{eff}}$ and  MET/$\sqrt{H_T}$ distinguish signal regions with the same jet multiplicity, ensuring a sensitivity to a range of mediator masses for each decay mode. In addition, seven signal regions target gluino pair production. These signal regions require either a minimum of four or five jets with different cuts on  $M_\text{eff}$ and MET. Sensitivity to a large range of mediator masses is ensured by cuts on $M_{\text{eff}}$ and MET/$M_\text{eff}(N_j)$.\footnote{$M_\text{eff}(N_j)$ is the scalar sum of the transverse momenta of the leading $N_j$ jets.} In both sets of signal regions, additional selection cuts are applied on jet momenta and pseudorapidities in order to reduce the SM background while maintaining high efficiencies. Moreover, in order to reduce the backgrounds from events with neutrinos ($W\to e\nu/\mu \nu$), events with isolated electrons or muons are rejected. The main backgrounds are:
\begin{itemize}
\item Z/$\gamma*$ + jets
\item W + jets
\item Diboson production
\item t$\bar{\text{t}}$(+EW) + single top
\end{itemize}
After applying all the cuts required by the search, the multi-jet background due to MET from misreconstructed jets in the calorimeter or jets misidentified as electrons is negligible.

We verified that we could reproduce their results with good accuracy for their SUSY signals. A point in the parameter space is rejected if the number of events in any of the signal regions is above its individual $95\%$ CLs upper limit \cite{Read:2002hq, Junk:1999kv}. A number of 50000 events were generated for each grid point. At low mediator mass, around $100$ GeV, the efficiencies are very small and a larger number of points has been generated when needed. 

\begin{figure}[t!]
  \centering
  \begin{subfigure}{0.48\textwidth}
    \centering
    \includegraphics[width=\textwidth, viewport = 0 0 450 366]{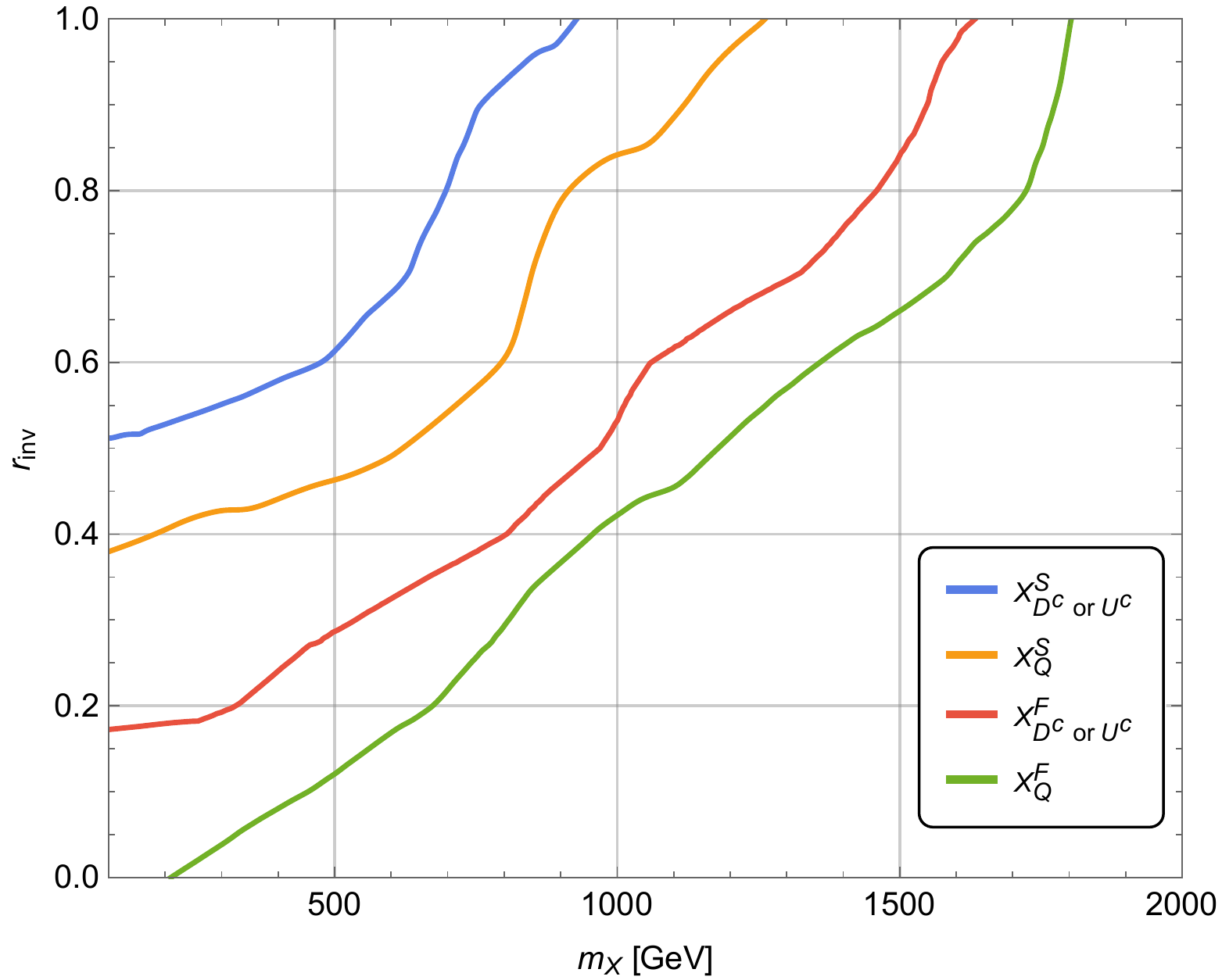}
    \caption{}
  \end{subfigure}
  ~
    \begin{subfigure}{0.48\textwidth}
    \centering
    \includegraphics[width=\textwidth, viewport = 0 0 450 366]{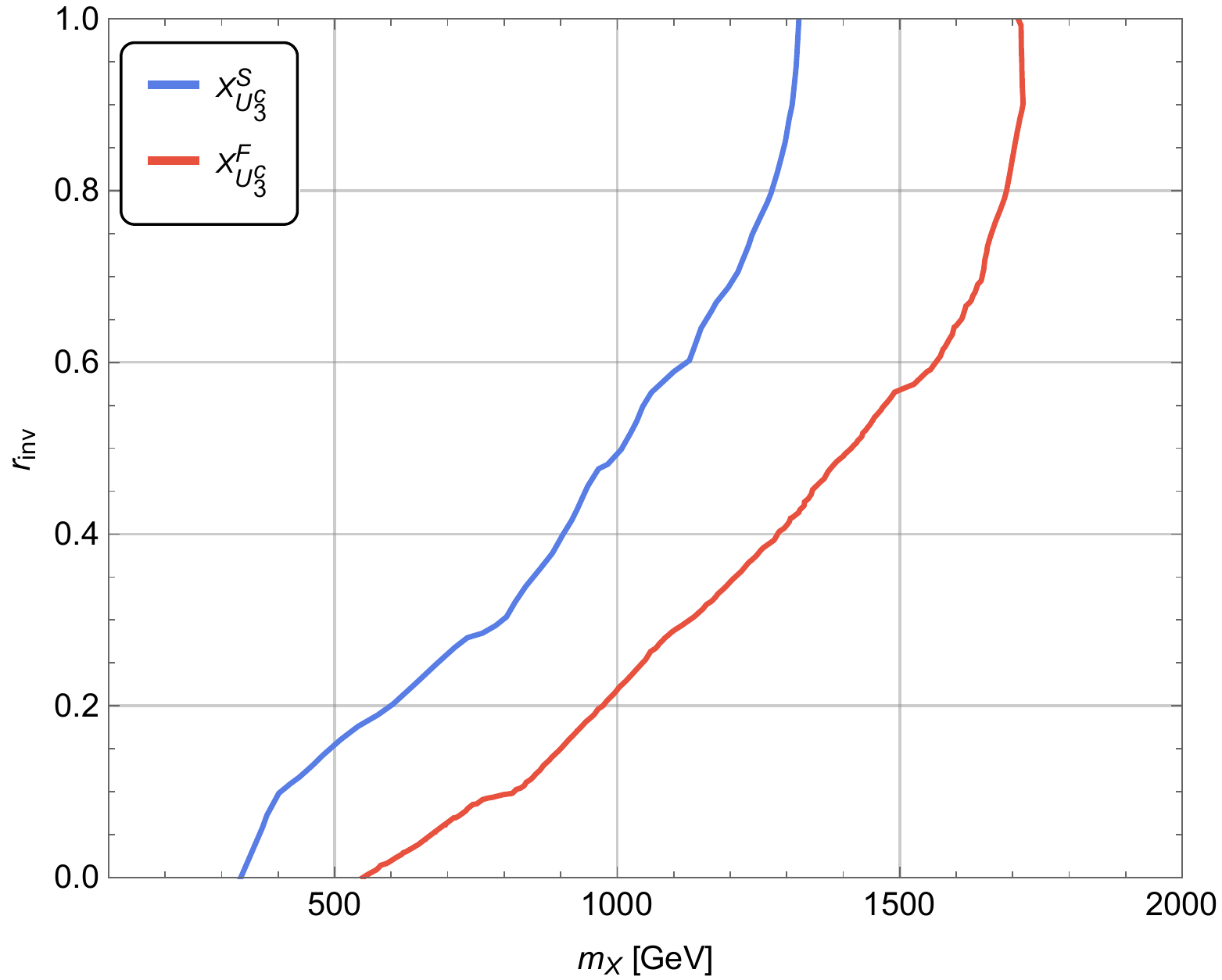}
    \caption{}
  \end{subfigure}
  ~
  \begin{subfigure}{0.48\textwidth}
    \centering
    \includegraphics[width=\textwidth, viewport = 0 0 450 366]{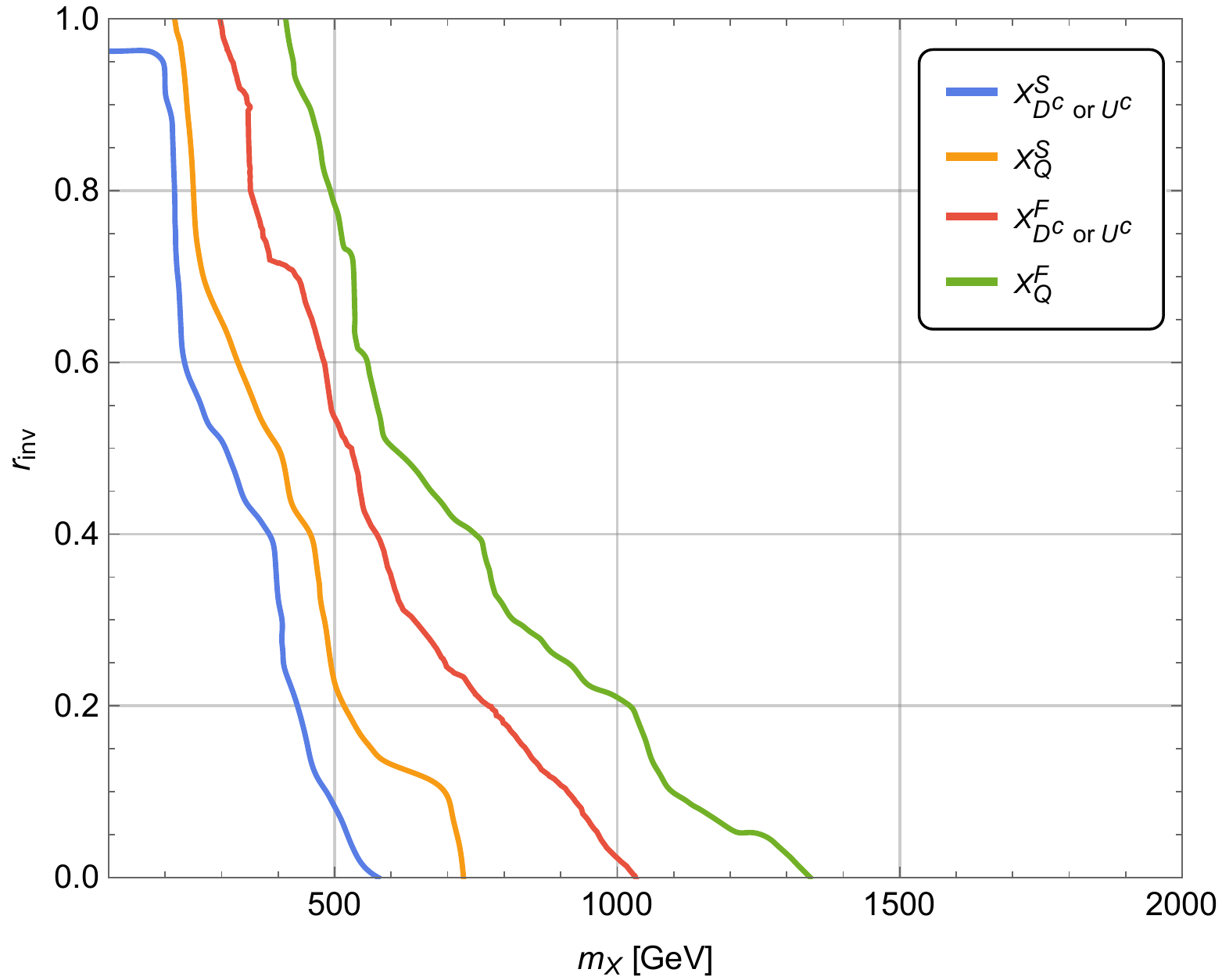}
    \caption{}
  \end{subfigure}
  ~
    \begin{subfigure}{0.48\textwidth}
    \centering
    \includegraphics[width=\textwidth, viewport = 0 0 450 366]{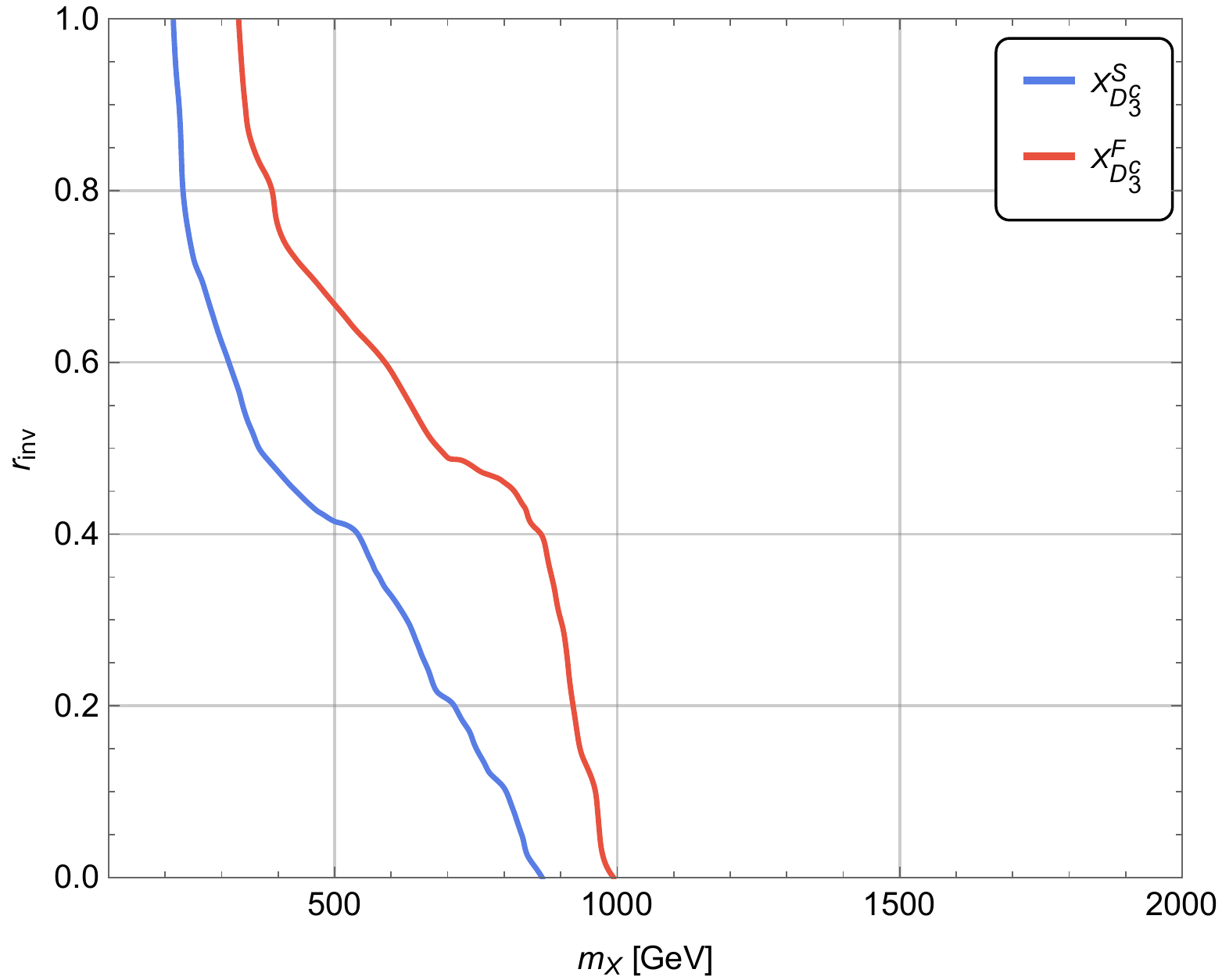}
    \caption{}
  \end{subfigure}

\caption{Limits at 95$\%$ CLs on mediators of category I that communicate only with (a) first and second generation quarks, (b) right-handed top quarks for searches targeting large $r_\text{inv}$ (left-handed top quarks give similar results), (c) first and second generation quarks and (d) left and right-handed bottom quarks for searches targeting small $r_\text{inv}$. Bounds are obtained by recasting the SUSY searches of Refs. (a) \cite{ATLAS:2017cjl}, (b) \cite{ATLAS:2016ljb}, (c) and (d) \cite{ATLAS:2017gsy}. The Hidden Valley group is taken to be $SU(3)$. The excluded regions are left of the curves.}\label{fig:CaseI_large_rinv}
\end{figure}

Results are presented in the top-left panel of Fig.~\ref{fig:CaseI_large_rinv} for $\mathcal{G}$ corresponding to $SU(3)$. Limits for mediators coupling to first and second generation quarks are the same. Stronger limits are obtained for fermion mediators because of their larger cross sections. Once $r_\text{inv}$ decreases, the limits weaken then disappear. This is due to the fact that for smaller $r_\text{inv}$ we expect more than two jets with large $p_T$ but an insufficient amount of MET. Relaxing the cut on MET could help to obtain stronger constraints for smaller $r_\text{inv}$, but the signal will have to compete with multi-jet background. Notice that for the mediator $X^F_{Q_{1,2}}$ we can exclude down to $r_\text{inv}=0$ for $m_X \lesssim 200$ GeV. This is due to the fact that the cross section is very large ($\mathcal{O}(10^{4})$ pb) and even the small probability of generating a large amount of MET from detector effects is enough to cause a sizable number of signal events. The bounds obtained for $r_\text{inv}=1$ are weaker with respect to the ones for squark production of Ref.~\cite{ATLAS:2017cjl}. This happens because they consider degenerate squarks for the first two generations, which results in their cross sections being much larger.

In the case of mediators coupling to third generation quarks, the constraints are stronger. If the mediator couples to the top, we can recast Ref.~\cite{ATLAS:2016ljb}. This analysis studies stop pair production with a final state of one isolated lepton, jets and MET. The data was taken at $\sqrt{s}=13$~TeV and corresponds to an integrated luminosity of 13.2 fb$^{-1}$. The pair produced stops decay to top and lightest neutralino. Finally, the tops decay to W bosons and b quarks, where one of the $W$ decays to an electron or muon and the $W$ from the other top quark decays hadronically. All events are required to have MET~$>$~200~GeV and need to have one signal lepton. Moreover, other event selection criteria suppress the multi-jet processes where leptons are misidentified or non-prompt and MET is mismeasured to a negligible level. The dominant backgrounds are $t\bar{t}$ events with at least one $W$ decaying leptonically and $W+$jets production. These backgrounds are suppressed by requiring the transverse mass of the signal lepton and the MET ($m_T$) to be larger than the $W$ mass. Furthermore, $t\bar{t}$ events where both $W$ decay leptonically or one leptonically and one with a hadronic $\tau$ decay are suppressed using variants of the transverse mass or the topness variable \cite{Graesser:2012qy}. 

Seven signal regions are then constructed using the discriminating variables just discussed. Two of these signal regions are used to cover stop production with subsequent decay $\tilde{t}\to t \tilde{\chi}^0$, giving the strongest constraints in our case. The first signal region (SR1) targets low mass splitting between the stop and the neutralino, with the decay products fully resolved. The second signal region ($\text{tN}\_\text{high}$) targets high mass splitting, leading to boosted top quarks. Therefore, the decay products are close and can be reconstructed within a single large-$R$ jet. We focus on these two signal regions because they reproduce better our signature and give the strongest limits. The main backgrounds for these signal regions are:
\begin{itemize}
\item t$\bar{\text{t}}$
\item W/Z + jets
\item single top
\item Diboson production
\item t$\bar{\text{t}}$ + V
\end{itemize}

We generate 50000 events and we use the public code \texttt{CheckMATE} \cite{Dercks:2016npn} to get the efficiencies for these operators. 

Results are presented in the top-right panel of Fig.~\ref{fig:CaseI_large_rinv} for $\mathcal{G}$ corresponding to $SU(3)$. Constraints for mediators interacting with left and right-handed tops give similar results. Note again that the limits for fermion mediators are stronger due to the larger cross sections. Limits on these mediators are stronger than the ones for mediators interacting only with first and second generation quarks, reaching both larger masses and lower $r_\text{inv}$. 

Mediators communicating only with bottom quarks are qualitatively similar to those communicating with the first two generations. The only difference is the possibility to require b-jets, which could potentially strengthen some of the bounds. We leave the exact details of this for future work.

\subsection{Case I with small $r_{\text{inv}}$}\label{sSec:CaseISmallrinv}
When $r_\text{inv}$ is small, the signature of the process changes considerably with respect to the previous case. In fact, we will have the following signatures:
\begin{enumerate}
\item 2 high $p_T$ jets/b-jets/tops, 2 low $p_T$ jets and MET
\item 2 high $p_T$ jets/b-jets/tops and 2 high $p_T$ jets
\end{enumerate}
The smaller $r_{\text{inv}}$, the larger is the fraction of dark hadrons that decay promptly to SM hadrons: in the limit of $r_{\text{inv}}=0$ all the dark hadrons decay promptly, leading to signatures with only jets and no MET. As $r_{\text{inv}}$ increases, the fraction of stable dark hadrons becomes larger, leading to signatures with 4 jets and some MET. 

Mediators of category I can be constrained by searches for paired dijets when $r_{\text{inv}}$ is small. A good example is the ATLAS search of Ref.~\cite{ATLAS:2017gsy}, which looks for pair produced colored resonances each decaying to two jets. Results were interpreted in models where a stop lightest supersymmetric particle decays via a baryon number breaking $R$-parity violating (RPV) coupling. Pair production of colorons decaying into two jets was also studied, considering an even larger energy range. The data was collected at $\sqrt{s}=$13~TeV with an integrated luminosity of $36.7$~fb$^{-1}$.

Events are counted as signal if they contain at least four reconstructed jets with $p_T>120$ GeV and $|\eta|<2.4$. This analysis aims at the case where the resonances are produced with large transverse momentum such that the decay products are expected to be close-by. Consequently, two jet pairs can be identified by considering every possible way to pair two jets, calculating a discriminating quantity that depends on the angular distance between the two jets of a given pair and then selecting the pairing that minimizes this quantity. Furthermore, a quality criterion on the pairing and an angular variable were defined to reduce the non-resonant multi-jet background. A mass asymmetry was then defined to discriminate between events with two equal mass reconstructed resonances. Dedicated signal regions with two b-tag jets are used for scenarios with couplings involving third generation quarks. Finally, the average mass of the two reconstructed resonances, $m_{\text{avg}}$, is used to discriminate the signal over a non-peaking background from multi-jet processes. 

The dominant backgrounds are from multi-jets events. Furthermore, the b-tagged signal regions have a non-negligible t$\bar{\text{t}}$ background.

The analysis includes a set of 40 overlapping signal regions that were optimized for different resonance masses. When applying constraints, we consider not only the bin optimized for the mass of the mediators but also all other bins. This is motivated by the fact that, looking only at this signal region, we would target only the case where $r_{\text{inv}}=0$. A non-zero $r_{\text{inv}}$ will shift $m_{\text{avg}}$ to lower values and therefore outside of its optimized bin. Therefore, we consider all 40 overlapping signal regions in terms of $m_{\text{avg}}$ windows, spanning from $100$ GeV to $2$ TeV. These signal regions are used to set limits on mediators coupling to first and second generation quarks. Mediators decaying to $n$ and b quarks can be better constrained using the two tag strategy. In addition to the requirements applied for the previous case, we require at least two b-tagged jets and impose that each of the b-tagged jets be associated with a different reconstructed resonance. This considerably reduces the backgrounds from multi-jet events. 

We verified that we could reproduce the results of Fig.~4 of Ref.~\cite{ATLAS:2017gsy} with good accuracy for their SUSY RPV signals. A point of parameter space is considered excluded if any individual signal region contains more events than its $95\%$ CLs upper limit. A minimum number of 50000 events were generated for each grid point. At low mediator mass, below $300$ GeV, the efficiencies are very small and a larger number of points has been generated when needed. 

Results are shown in Fig.~\ref{fig:CaseI_large_rinv} for $\mathcal{G}$ corresponding to $SU(3)$. The bottom-left panel corresponds to limits for mediators coupling to first and second generation quarks, while the bottom-right one shows constraints for mediators coupling to bottom quarks. Constraints for mediators interacting with left and right-handed bottoms give the same result. As expected from Fig.~\ref{fig:CaseI_CS_NLO}, the fermion mediators lead to stronger constraints due to their larger cross sections. This search is stronger at $r_\text{inv}=0$, degrading slowly as $r_\text{inv}$ becomes larger. This is also expected because the reconstructed mass of dijets goes as $\sqrt{1-r_\text{inv}}m_X$ and therefore a sizable $r_\text{inv}$ is needed to suppress enough the dijet mass. These bounds are unsurprising as light new colored states would be extremely hard to miss at the LHC. 

Do note that paired dijet searches would start to lose their effectiveness if semivisible jets had a large variation on the fraction of their energy that goes to invisible. This would be the case for a large confinement scale. Also, similar results could be obtained for mediators decaying to tops, but paired-dijet searches would probably not be the optimal search strategy in that case.

A possible way to improve this search for intermediate values of $r_{\text{inv}}$ is to require a cut on MET. Indeed, for $r_\text{inv} > 0$, we expect an amount of  MET that increases as $r_{\text{inv}}$ increases. As a benefit, a cut on MET would considerably reduce the QCD background and reduce the previously absent Z+jets background, where we expect the MET distribution to peak at small values. 

\section{Case II: Three field operators involving leptons}\label{Sec:CaseII}
Case II consists of three field operators involving the mediator, a dark (s)quark and a lepton. At hadron colliders, the mediators $X$ will mainly be pair produced via an $s$-channel photon, $Z$ or possibly $W$. Depending on which generations $X$ couples with, pair production could also be possible at lepton colliders via diagrams involving a $t$-channel $n$. Once produced, the mediator will decay to a semivisible jet and either a lepton, a tau or a neutrino.\footnote{Following the experimental convention, we refer from now on to leptons as only electrons and muons.} In general, the experimental signatures of case II can be classified into three distinct categories:
\begin{enumerate}
  \item 2 charged leptons/taus, 2 semivisible jets and MET
  \item 1 charged lepton/tau, 2 semivisible jets and more MET
  \item 0 charged leptons/taus, 2 semivisible jets and even more MET
\end{enumerate}
Of course, point 2 and 3 are only possible if the mediator is not an $SU(2)_L$ singlet. If the full theory only contained mediators of category II, dark hadrons would typically decay to leptons, leading to lepton jets instead. It is possible for mediators of category II to still lead to semivisible jets as long as a mediator from another category is also present. We simply assume that this other mediator is too heavy to be kinematically accessible, but that its coefficient $\lambda$ is much higher than the ones from category II. 

We now discuss the cases of large and small $r_{\text{inv}}$ which can be constrained by current searches. We then explain how these search strategies would need to be modified to account for intermediary values of $r_{\text{inv}}$.

\subsection{Case II with large $r_{\text{inv}}$}\label{sSec:CaseIILargerinv}
In the limit of $r_{\text{inv}} \to 1$, semivisible jets transmit all of their energy to MET. The signatures then become equivalent to those of $R$-parity conserving Supersymmetry. More precisely, they resemble pair produced sleptons decaying to a stable neutralino and a charged lepton, tau or neutrino. For sleptons, the signature that is most commonly looked for at colliders is MET and two leptons or taus.

To obtain collider bounds for the first two generations and large $r_{\text{inv}}$, we recast the ATLAS search of Ref.~\cite{ATLAS:2017uun}. The data was collected at $\sqrt{s}=13$~TeV with 36.1 fb$^{-1}$ of integrated luminosity. More specifically, we concentrate on the signal regions for 2 same flavor leptons and 0 jets. In addition to basic cuts on the reconstructed leptons, the signal regions consist of cuts on the invariant mass of the two leptons $m_{ll}$ and the stransverse mass $M_{T2}$ \cite{Lester:1999tx, Barr:2003rg}. A veto is placed on jets that were not b-tagged with a transverse momentum of $p_T > 60$~GeV or b-tagged jets with $p_T > 30$~GeV. We verified that we could reproduce their results with good accuracy for their SUSY signals. Finally, we reject a point of parameter space if the number of events in any of the two inclusive signal regions is above its individual 95$\%$ CLs upper limit. A number of 20000 events were generated for each grid point. The main backgrounds are:
\begin{itemize}
  \item Diboson production
  \item Drell-Yan + jets
  \item $t\bar{t}$ production
  \item $Wt$ production
  \item Fake leptons.
\end{itemize}

\begin{figure}[t]
\begin{center}
    \includegraphics[width=0.6\textwidth, viewport = 0 0 450 375]{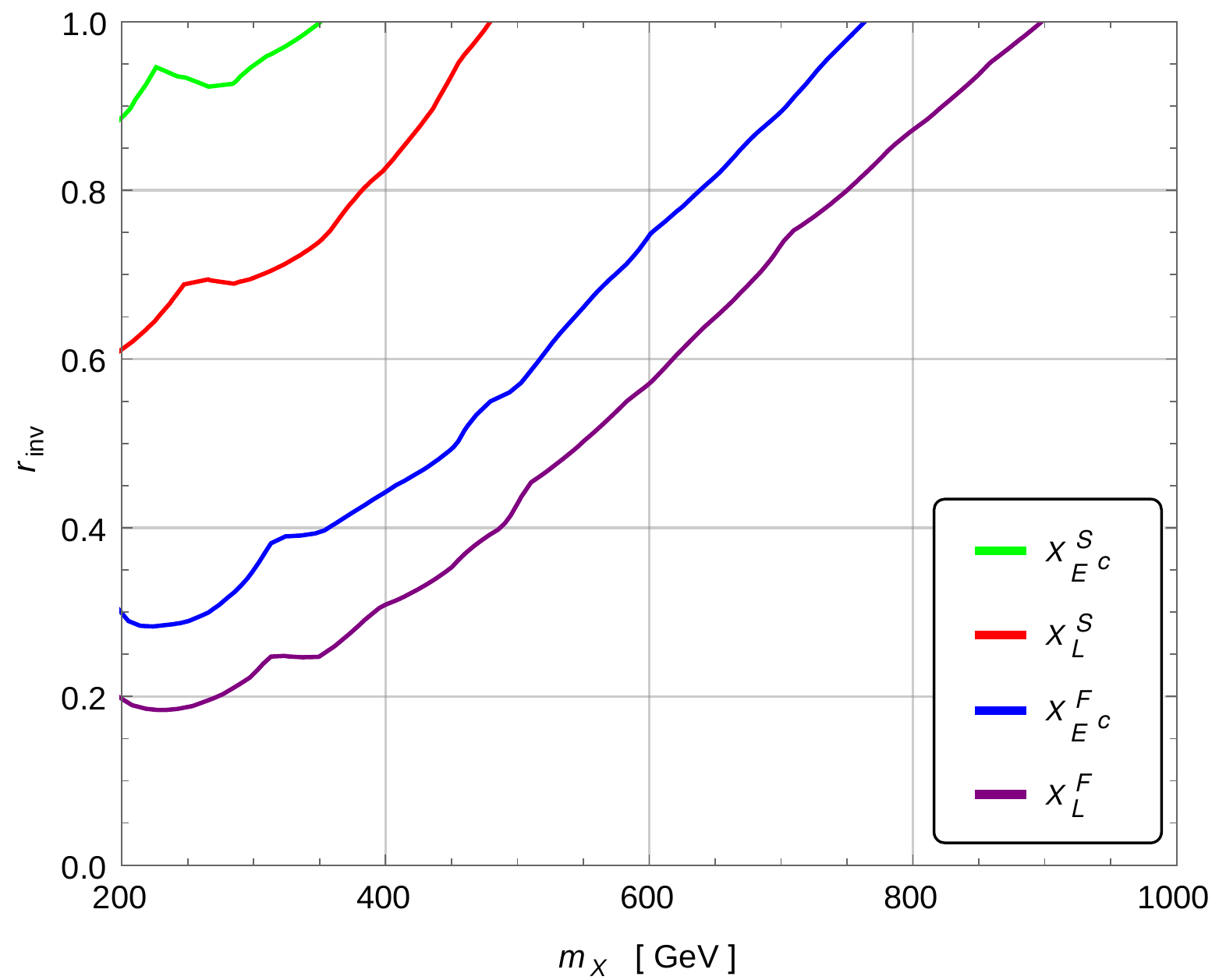}
  \end{center} 
\caption{Limits at 95\% CLs on mediators of category II that communicate only with the first generation for large $r_{\text{inv}}$. Bounds are obtained by recasting the slepton search of Ref.~\cite{ATLAS:2017uun}. The Hidden Valley group $\mathcal{G}$ is taken to be $SU(3)$. The excluded regions are left of the curves.}\label{fig:CaseIILargeRinv}
\end{figure}

Results are presented in Fig. \ref{fig:CaseIILargeRinv} for $\mathcal{G}$ corresponding to $SU(3)$ and mediators communicating with the first generation only. 
Limits for the second generation are very similar, albeit slightly stronger because of a larger muon reconstruction efficiency. Constraints for fermion mediators are considerably stronger simply because of a much higher cross section. As was to be expected, the limits drop spectacularly once $r_{\text{inv}}$ starts to decrease. This is even more accentuated because of the cut on jets. For scalar mediators, this is not much of a problem as the jets are typically not too energetic. This is more of a problem for fermion mediators, which can be probed at much higher masses where semivisible jets are much more energetic.

\begin{figure}[t!]
  \centering
  \begin{subfigure}{0.48\textwidth}
    \centering
    \includegraphics[width=\textwidth, viewport = 0 0 450 366]{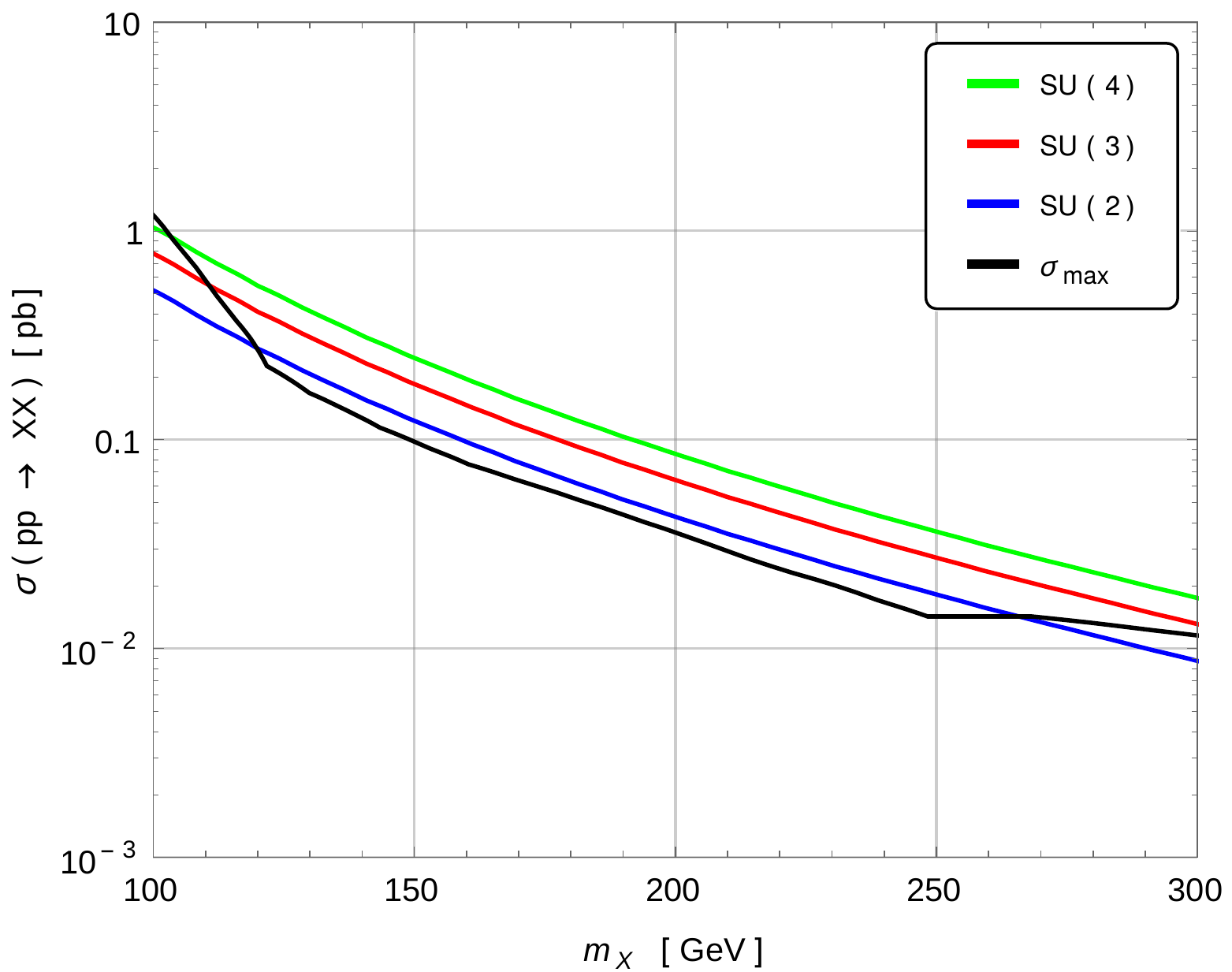}
    \caption{}
  \end{subfigure}
  ~
    \begin{subfigure}{0.48\textwidth}
    \centering
    \includegraphics[width=\textwidth, viewport = 0 0 450 366]{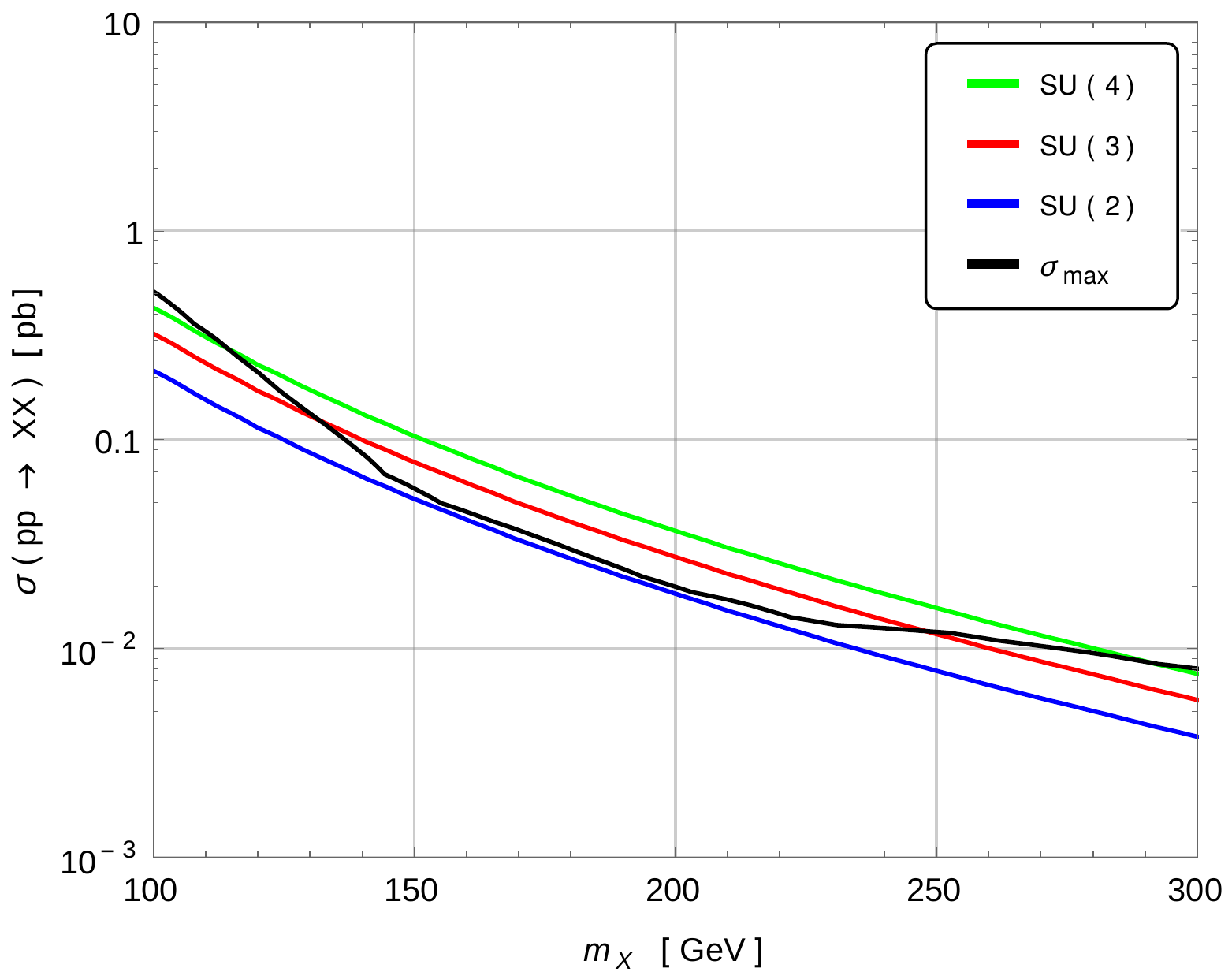}
    \caption{}
  \end{subfigure}
\caption{Upper limits at 95\% CLs on the cross section of pair production of (a) $X^S_L$ and (b) $X^S_{E^c}$ from the LHC stau search of Ref.~\cite{CMS:2017rio} and $r_{\text{inv}}\to 1$. Cross sections are also presented for $\mathcal{G}$ corresponding to different $SU(N)$.}\label{fig:ThirdGenerationStau}
\end{figure}

Similar results can be obtained for operators that communicate with the third generation instead. Ref.~\cite{CMS:2017rio} consists of a similar search but with leptons replaced by hadronically decaying taus. The data was taken at $\sqrt{s}=13$~TeV and corresponds to an integrated luminosity of 35.9 fb$^{-1}$. It requires the presence of two hadronically decaying taus. Besides basic cuts on the reconstructed tau jets, the kinematic cuts are on the stranverse mass, the MET, the sum of the transverse mass of the two tau jets and the difference between their azimuthal angles. A veto is applied on other light leptons and b-tagged jets. In this case, a limit on the maximal cross section is included. This allows us to put limits on scalar mediators in the limit of $r_{\text{inv}}\to 1$ without any simulations. Unfortunately, the search is not very effective at low masses and generally no lower limits can be put on scalar mediators. In Fig.~\ref{fig:ThirdGenerationStau}, we show the upper limit on the cross section $\sigma_{\text{max}}$ for left-handed and right-handed scalar mediators. These limits cannot be applied directly to fermion mediators and we leave the task of obtaining limits in this case for future work. The results are not expected to be qualitatively different from those for first and second generations.

\subsection{Case II with small $r_{\text{inv}}$}\label{sSec:CaseIISmallrinv}
Case II with a small $r_{\text{inv}}$ represents a much greater challenge. The different collider signatures are akin to those of large $r_{\text{inv}}$, but with more energetic jets and far less MET. In the limit of $r_{\text{inv}}$ going to zero, these signatures become identical to leptoquark pair production. The main difference is that the mediators are produced via an electroweak process as opposed to a strong one. This greatly reduces the cross sections and renders most search strategies far less effective.

To illustrate the situation, we recast the leptoquark search of Ref.~\cite{CMS:2016imw}. The data was taken at $\sqrt{s}=13$~TeV with an integrated luminosity of 2.6 fb$^{-1}$. Although the integrated luminosity is rather low, this search has the advantage of containing signal regions that are optimized for small masses. Typically, leptoquarks of a few hundred GeV have been ruled out several years ago and many searches only contain signal regions that have been optimized for much heavier ones. Though this is certainly a valid approach for these searches, it generally results in kinematic cuts that fail badly for lighter leptoquarks. Since the cross section for pair production of mediators of case II is much smaller, the limits would be expected to fall in these regions, if they even exist at all. This makes many searches ineffective for constraining case II, but not the one of Ref.~\cite{CMS:2016imw} though.

To be counted as a signal, an event is required to contain exactly two electrons and at least two jets. Besides a few basic cuts on reconstructed objects, the kinematic cuts of the different signal regions are based on three different quantities. The first one is $S_T$, defined as the scalar sum of the $p_T$ of the leptons and the two leading jets. The second is the invariant mass of the two leptons. The third is $m_{ej}^{min}$, defined by selecting the jet-electron pairing that minimizes the invariant mass difference and taking the smallest invariant mass of the two. A set of 27 overlapping signal regions are then defined in terms of these variables, each signal region being optimized for a different leptoquark mass. We exclude a point if any individual signal region contains more signals than its 95$\%$ CLs upper limit. We verified that we could reproduce their leptoquark results with good accuracy. A number of 20000 events were generated. The main backgrounds are:
\begin{itemize}
  \item $Z$ + jets
  \item $t\bar{t}$
  \item QCD.
\end{itemize}

\begin{figure}[t!]
  \centering
  \begin{subfigure}{0.48\textwidth}
    \centering
    \includegraphics[width=\textwidth, viewport = 0 0 450 366]{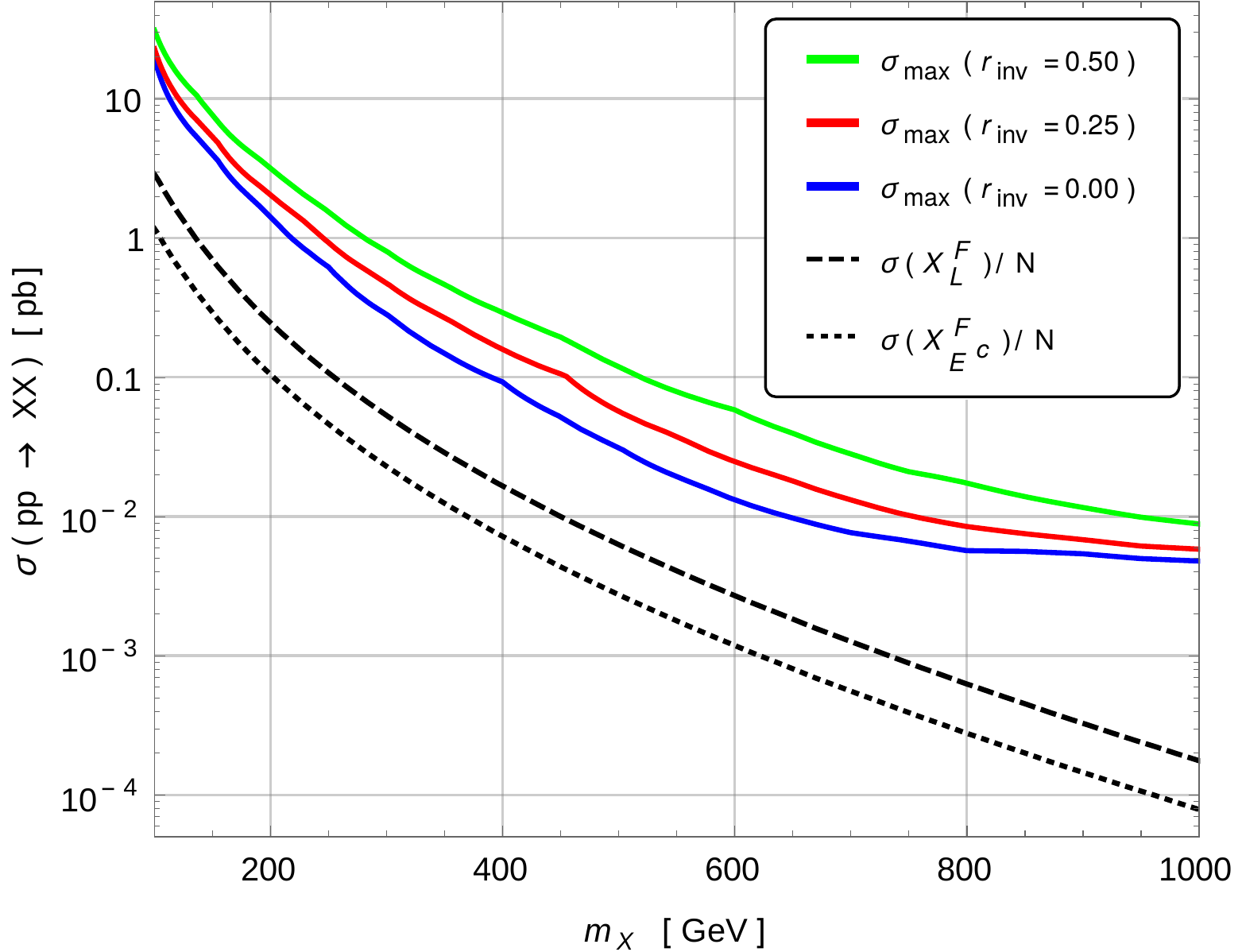}
    \caption{}
  \end{subfigure}
  ~
  \begin{subfigure}{0.48\textwidth}
    \centering
    \includegraphics[width=\textwidth, viewport = 0 0 450 366]{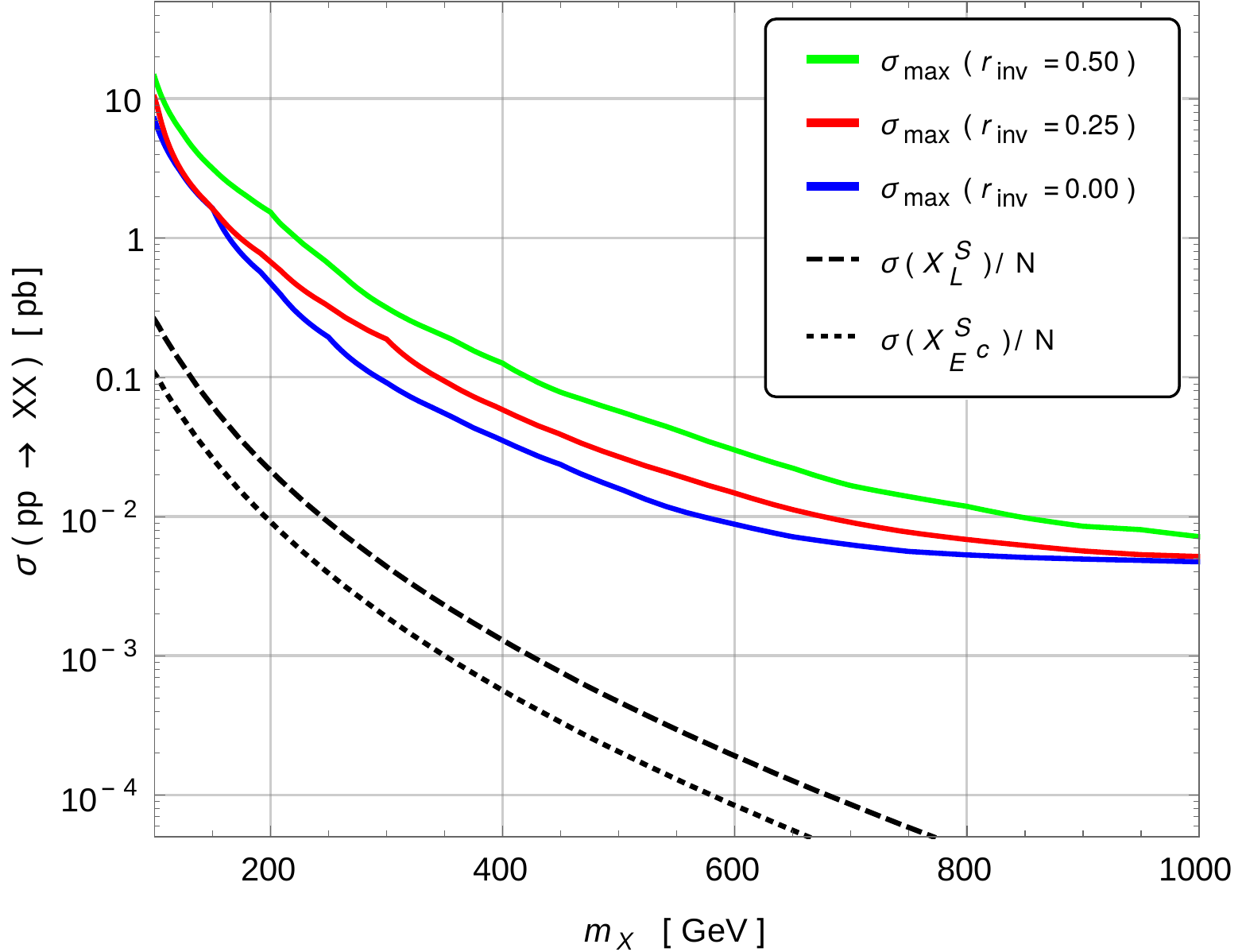}
    \caption{}
  \end{subfigure}
\caption{Upper limits at 95\% CLs on the cross section for (a) fermion mediators and (b) scalar mediators of category II from the LHC leptoquark search of Ref.~\cite{CMS:2016imw} and $r_{\text{inv}}\to 0$. Assuming $\mathcal{G}$ to be $SU(N)$, the different cross sections divided by $N$ are also presented.}\label{fig:LHCleptoquarks}
\end{figure}

Fig.~\ref{fig:LHCleptoquarks} shows the excluded cross section at 95$\%$ confidence level as a function of the mass of the mediator for several small values of $r_\text{inv}$. Due to their nature as either vector fermions or scalars, the maximal cross section is unaffected by whether the mediator couples to left-handed or right-handed electrons. The bounds were obtained for $\mathcal{G}$ corresponding to $SU(3)$, but the maximal cross section does not vary much for other groups. Assuming $\mathcal{G}$ to be $SU(N)$, the cross section per $N$ is also shown. No constraints are obtained for fermion mediators with $N$ inferior to 5 or 10 for $X^F_L$ and $X^F_{E^c}$ respectively. Similarly, no constraints are obtained for scalar mediators with $N$ inferior to about 23 or 55 for $X^S_L$ and $X^S_{E^c}$ respectively. Similar results are expected for mediators interacting with muons, but the bounds for those coupling with taus would typically be weaker due to these particles leading to less clean signatures.

Despite generally not providing any constraints, searches strategies like this one leave little room for improvements. This is a general feature of hadron colliders, as many of the main backgrounds can be generated by strong processes while the mediators are electroweakly produced. Lepton colliders do not share this problematic and represent probably the best opportunity to discover mediators of category II for small $r_{\text{inv}}$. 

Bounds can be extracted from LEP data in the limit of small $r_{\text{inv}}\to 0$. This is done by recasting the search of Ref.~\cite{Abbiendi:2003iv} for leptoquark pair production. This analysis combines data taken at energies varying between 189 and 208~GeV, representing a total integrated luminosity of 595.9~pb$^{-1}$. The produced leptoquarks can either both decay to charged leptons, both decay to neutrinos or one to a charged lepton and the other to a neutrino. The leptoquarks are assumed to communicate with only a single generation and every lepton generation is considered. Both scalar and vector leptoquarks are studied and efficiencies tables are provided in both cases. Since the efficiencies for scalar and vector leptoquarks are very similar, we neglect the effect of the spin of the leptoquark. This approximation was also made in Ref.~\cite{Diaz:2017lit} and the difference between the efficiencies was found to be generally below 10\%. We then create a set of maps of the efficiency as a function of the mass of the leptoquark and the center of mass energy for all signal regions. Calculating cross sections once again with \texttt{MadGraph5\_aMC@NLO}, the number of signals can be obtained for every signal region. The model is considered excluded if a single signal region contains more signals than its 95\% CLs upper limit. Lower limits are shown in table~\ref{tab:CaseIILEPlimits} for $\mathcal{G}$ corresponding to $SU(N)$ and for $X$ communicating with any single generation. Unsurprisingly, all limits are close to the maximal kinematic reach of LEP. This is simply a consequence that bounds for leptoquarks were already very close to their maximal value and that the mediators usually have a much higher cross section. Finally, note that LEP bounds on single production of leptoquarks do not apply.

\begin{table}[t]
{\footnotesize
\setlength\tabcolsep{5pt}
\begin{center}
\begin{tabular}{ccccc}
\toprule
Mediator    & Generation & $SU(2)$ [GeV] & $SU(3)$ [GeV] & $SU(4)$ [GeV] \\
\cmrule
            & 1          & 104           & 104           & 104     \\
$X^F_{L}$   & 2          & 104           & 104           & 104     \\
            & 3          & 103           & 104           & 104     \\
            & 1          & 103           & 104           & 104     \\
$X^F_{E^c}$ & 2          & 104           & 104           & 104     \\
            & 3          & 103           & 103           & 104     \\
            & 1          & 93            & 95            & 96      \\
$X^S_{L}$   & 2          & 98            & 99            & 100     \\
            & 3          & 91            & 93            & 95      \\
            & 1          & 93            & 95            & 96      \\
$X^S_{E^c}$ & 2          & 98            & 99            & 100     \\
            & 3          & 90            & 93            & 94      \\
\bottomrule
\end{tabular}
\end{center}
}
\caption{Lower limit on the mass of mediators of category II at 95\% CLs for $r_{\text{inv}}=0$. Generation corresponds to the single generation of leptons the mediator couples with. Bounds are given for $\mathcal{G}$ corresponding to different $SU(N)$ and originate from the LEP search of Ref.~\cite{Abbiendi:2003iv} for leptoquarks.} 
\label{tab:CaseIILEPlimits}
\end{table}

\subsection{Case II with intermediary $r_{\text{inv}}$}\label{sSec:CaseIIMediumrinv}
An intermediary value of $r_{\text{inv}}$ represents the scenario that is most distant from any current search strategies. As such, there are no present experimental analysis that can give optimized bounds on the mediators. In this context, we limit ourselves to explaining how previous searches could have been modified to better constrain this region of parameter space. More precisely, we discuss how the slepton search of Ref.~\cite{ATLAS:2017uun} would need to be modified to more efficiently constrain intermediary values of $r_{\text{inv}}$.

The cuts of Ref.~\cite{ATLAS:2017uun} failed for intermediary $r_{\text{inv}}$ partially because of the presence of a jet veto. Instead of this cut, one could require the presence of at least two jets above a threshold $p_T$, which we take for illustration purposes to be 20~GeV. The search of Ref.~\cite{ATLAS:2017uun} and the leptoquark search of Ref.~\cite{CMS:2016imw} then differ qualitatively only by Ref.~\cite{ATLAS:2017uun} requiring some MET and Ref.~\cite{CMS:2016imw} cutting on the momentum of the jets. The main background is $t\bar{t}$, as it was the second most important background despite being considerably suppressed by the veto on jets and the even stronger veto on b-jets. We can then pair every lepton with a jet and select the pairing that minimizes the difference between the two invariant masses, just like in Ref.~\cite{CMS:2016imw}. In an ideal scenario, this would always select the two jets that originate from the mediator decay. The two lepton-jet pairs should then reconstruct to an invariant mass of about $\sqrt{1-r_{\text{inv}}}m_X$, where $m_X$ is the mass of the mediator. Unfortunately, $m_{ej}^{min}$ is not able to discriminate efficiently between the signal and $t\bar{t}$. Since $m_X$ is not expected to be more than a few hundred GeV, $m_{ej}^{min}$ should not be much higher than a couple of hundred GeV for intermediary values of $r_{\text{inv}}$. The distribution of $m_{ej}^{min}$ for the signal is then not so different than for the $t\bar{t}$ background, which itself peaks at around 175~GeV. However, the vectorial sum of the $p_T$ of the two jets originating from the mediators, which we label $p_T^{\text{X jets}}$, should point in the same direction as the MET. One can then define $\Delta\phi_{\text{MET}}^J$ as the angular difference between $p_T^{\text{X jets}}$ and the MET. This quantity turns out to be efficient at distinguishing the signal from other backgrounds.

\begin{figure}[t]
\begin{center}
    \includegraphics[width=0.6\textwidth, viewport = 0 0 450 400]{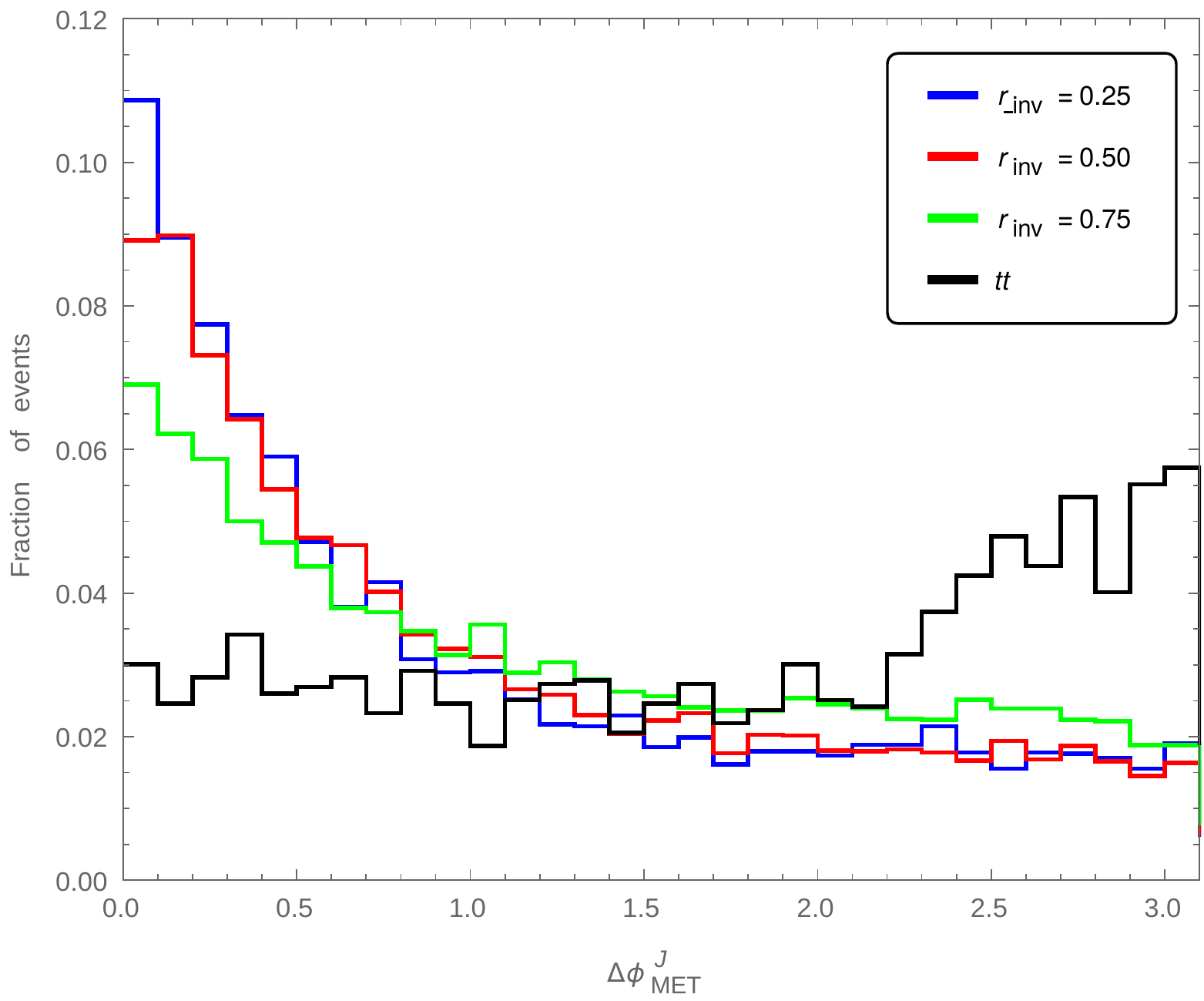}
  \end{center} 
\caption{Normalized distribution of $\Delta\phi_{\text{MET}}^J$ in case II for different signals and the $t\bar{t}$ background. The mediator was taken to be $X_{L}^S$ with a mass of 300~GeV. The cuts of Ref.~\cite{ATLAS:2017uun} for the lowest inclusive signal region for 2 same flavor leptons and 0 jets were applied beforehand, except the veto on jets and b-jets.} \label{fig:New_cut_1}
\end{figure}

As an illustration, we produced distributions of $\Delta\phi_{\text{MET}}^J$ for various mediators using the same procedure as before and generating $10^5$ events. We also generated 5 million $t\bar{t}$ backgrounds to serve as a comparison. To improve statistics, we required the tops to decay to electrons at \texttt{MadGraph} level. For both signals and backgrounds, the cuts from the lowest inclusive regions of Ref.~\cite{ATLAS:2017uun} were applied beforehand, except of course the veto on jets and b-jets. Fig.~\ref{fig:New_cut_1} shows the distribution of $\Delta\phi_{\text{MET}}^J$ for example signals and $t\bar{t}$. As can be seen, a cut on $\Delta\phi_{\text{MET}}^J$ could significantly reduce the background while leaving the signal mostly unscathed. Such a cut would obviously not be effective when $r_{\text{inv}}$ is too close to either 0 or 1, as the former would lack the MET and the later the proper jets. We also note that the fact that the $t\bar{t}$ background peaks around $\pi$ is due to the implicit requirement of a large amount of MET. This forces the neutrinos to be aligned in the same direction and the jets in the opposite one. The semivisible jet candidates are then most likely to be in the exact opposite direction of the MET, which make the cut on $r_{\text{inv}}$ even more effective. This effect is accentuated when the MET cut is made more stringent. The cut on the stransverse mass also affects the distribution of $\Delta\phi_{\text{MET}}^J$ to a certain extent. We leave the details of the optimization for future work. Of course, the $t\bar{t}$ background could be further reduced by applying a veto on b-tagged jets.

\section{Case III: Operators involving fermion mediators and the Higgs}\label{Sec:CaseIII}
Mediators of category III are three field operators involving the Higgs doublet with fermion $X$ and $n$. It was already studied in large part in Ref.~\cite{Beauchesne:2017ukw}. Only values of $r_{\text{inv}}$ of 0 or 1 were considered however. We follow a similar procedure but include experimental results that were published since then. The relevant Lagrangian is given by:
\begin{equation}\label{eq:LagrangianCaseIII}
  \mathcal{L} = -m_X \bar{X}^F_H X^F_H -m_n \bar{n}^Fn^F + \left(\lambda_{H,L}^F H^\dagger \bar{n}^F P_L X^F_H + \lambda_{H,R}^F H^\dagger \bar{n}^F P_R X^F_H + \text{h.c.}\right).
\end{equation}
Once the Higgs acquires an expectation value, the neutral component of $X_H^F$ mixes with $n^F$. The resulting spectrum is a charged fermion $C^+$ of mass $m_X$ and two neutral Dirac fermions. The heaviest one is labelled $N_2$ and is typically close in mass to $C^+$. The other neutral fermion is referred to as $N_1$ and is generally light. All of these fermions are still fundamental of the unbroken $\mathcal{G}$.

The charged and neutral fermions can be pair produced via diagrams involving an $s$-channel photon, $W$ or $Z$. Any combination of either $N_1$, $N_2$ or $C^+$ with any of their antiparticle can be produced this way. Once created, $N_2$ and $C^+$ will decay while $N_1$ remains stable. The charged fermion $C^+$ generally decays via $C^+ \to N_1 W^+$. The neutral fermion $N_2$ can decay via both $N_2 \to N_1 Z$ and $N_2 \to N_1 h$, though the former decay typically dominates. The decay $N_2 \to C^+ W^-$ is also possible, but it is phase-space suppressed and negligible for all sensible regions of parameter space. The possible signatures are then:
\begin{enumerate}
  \item 2 semivisible jets and MET
  \item 2 semivisible jets, MET and 1 $W$/$Z$/$h$
  \item 2 semivisible jets, MET and 2 $W$/$Z$/$h$
\end{enumerate}

\begin{figure}[t]
\begin{center}
    \includegraphics[width=0.6\textwidth, viewport = 0 0 450 400]{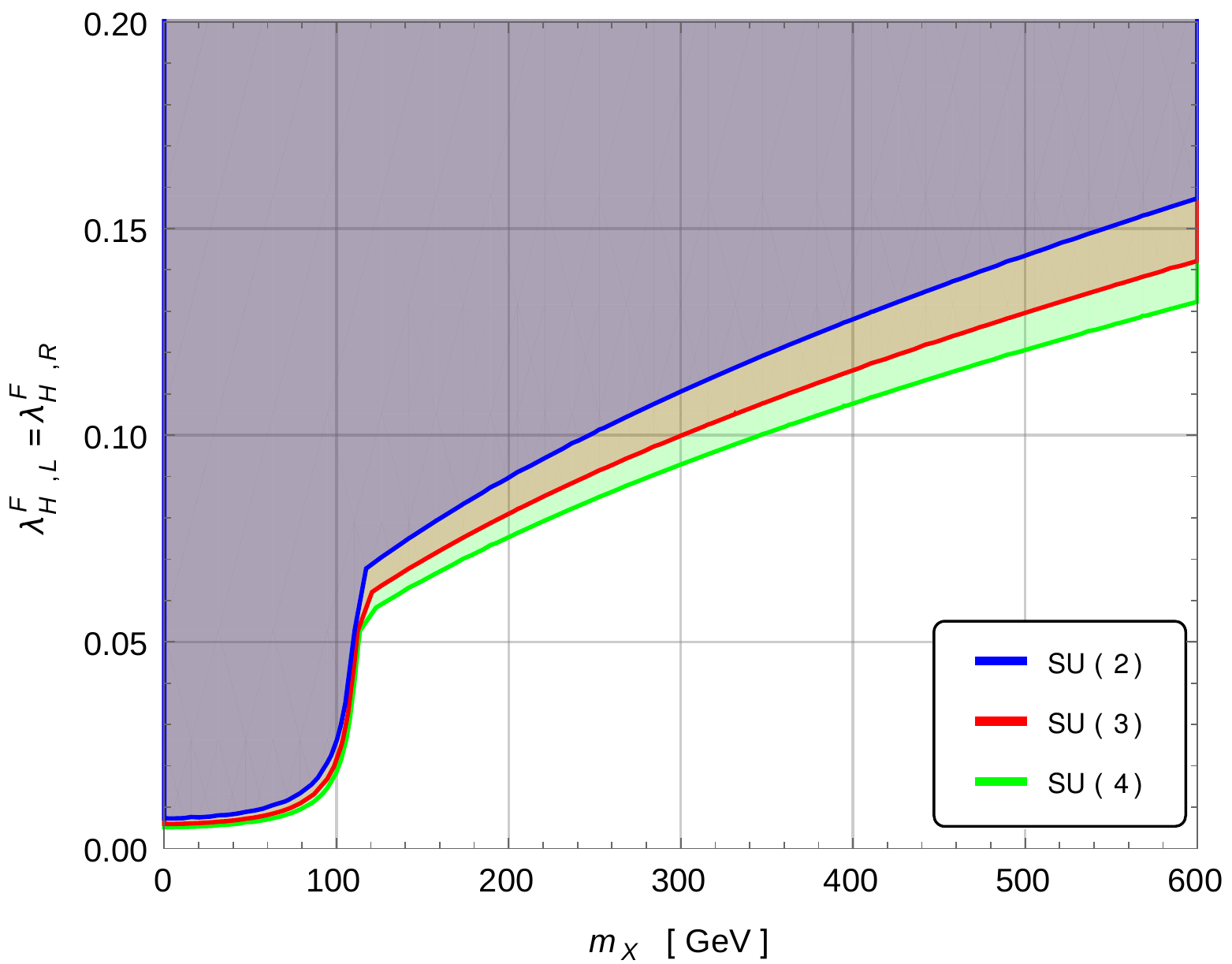}
  \end{center} 
\caption{Excluded region at 95\% confidence level in the ($m_X$, $\lambda_{H,L}^F=\lambda_{H,R}^F$) plane from the Higgs branching ratio to semivisible jets for different $\mathcal{G}$. The mass $m_N$ was set to 10~GeV.}\label{fig:CaseIIIHiggsyLyR}
\end{figure}

The parameters $\lambda_{H,L}^F$ and $\lambda_{H,R}^F$ are constrained by studies of the Higgs properties. The dominant effect is the tree-level decay of the Higgs to semivisible jets. These represent so-called undetectable objects \cite{Bechtle:2014ewa} and lead to a diminution of the branching ratios to well studied channels. A bound of 20\% on Higgs branching ratio to undetectable objects at 95\% confidence level was found in Ref.~\cite{Bechtle:2014ewa} for standard couplings of the Higgs to SM particles. The bound on the branching ratio to invisible is slightly stronger, but we neglect the difference. Loop effects can be neglected in first approximation. Constraints on $\lambda_{H,L}^F$ and $\lambda_{H,R}^F$ are obtained by requiring the Higgs branching ratio to all combinations of $\bar{N}_iN_j$ to be below this bound. Limits as a function of $m_X$ are shown in Fig.~\ref{fig:CaseIIIHiggsyLyR} by setting $\lambda_{H,L}^F=\lambda_{H,R}^F$. As can be seen, only small values of $\lambda_{H,L}^F$ and $\lambda_{H,R}^F$ are allowed. This result forces $C^+$ and $N_2$ to be almost degenerate in mass. We will exploit this fact in Sec. \ref{sSec:CaseIIIMediumrinv}.

We now discuss how current searches can constrain small and large values of $r_{\text{inv}}$. We then discuss how intermediary values could be better probed.

\subsection{Case III with  large $r_{\text{inv}}$}\label{sSec:CaseIIILargerinv}
In the limit of $r_{\text{inv}}\to 1$, the phenomenology of case III becomes similar to that of electroweakinos. The channel we focus on is the production of $N_2 C^-$ and its conjugate process. Of all the possible combinations of particles of mass around $m_X$ that can be pair produced, this is typically the one with the largest cross section. We concentrate on the decay $N_2\to N_1 Z$ as it is generally the dominant one and also the cleanest. The resulting signature is then a $W$, a $Z$ and MET. It is equivalent to the pair production of a neutralino and a chargino, with the neutralino decaying to a $Z$ and a lighter stable neutralino and the chargino decaying to a $W$ and the stable neutralino.

We obtain bounds in the limit of large $r_{\text{inv}}$ by recasting the electroweakino search of Ref.~\cite{Sirunyan:2017qaj}. The data was taken at $\sqrt{s}=13$~TeV and corresponds to an integrated luminosity of 35.9 fb$^{-1}$. More precisely, we focus on their VZ signal regions which are designed for leptonically decaying $Z$ and hadronically decaying $W$. Besides a few basic requirements on the reconstructed jets, the main cuts are as follow. The event is required to contain exactly two leptons. These are further required to be of opposite sign, have the same flavor and to have an invariant mass between 86 and 96~GeV. The stransverse mass is required to be above 80~GeV. A minimal number of 2 jets is required, but events containing b-tagged jets are vetoed. The two jets closest in terms of azimuthal angles are required to have an invariant mass inferior to 110~GeV. The two jets with the highest $p_T$ are required to have an azimuthal angle separation of at least 0.4 from the MET. Events are further classified into four signal regions depending on their MET. We verified that we could reproduce their efficiencies with good accuracy. The model is considered excluded if any bin contains more signals than its 95\% CLs upper limit. A number of 20000 events were generated for each simulated point of parameter space. The $Z$ was decayed leptonically and the $W$ hadronically at \texttt{MadGraph} level to improve statistics. The main backgrounds are:
\begin{itemize}
  \item Drell-Yan + jets
  \item Flavor symmetric backgrounds ($t\bar{t}$, $W^+W^-$...)
  \item $Z+\nu$
\end{itemize}

Unfortunately, the search of Ref.~\cite{Sirunyan:2017qaj} contains a sizable excess in one of its bins optimized for compressed spectra. This results in the upper limit on $r_{\text{inv}}$ being less than expected for intermediary masses. To address this issue, we also recast the search of Ref.~\cite{Sirunyan:2017lae}, which is similar but instead studies leptonic decay of the $W$. The data was taken at $\sqrt{s} = 13$~ TeV with 35.9~fb$^{-1}$ of integrated luminosity. No analogous excess is present in this search. More precisely, we concentrate on their three light-flavor leptons signal regions. As the name suggests, these signals regions require the presence of three leptons, two of which must additionally be of opposite sign and same flavor (OSSF). After a few basic kinematic cuts, events are binned into 44 signal regions according to their MET and the invariant mass of the two OSSF leptons. A veto is placed on b-tagged jets. We again generate 20000 events and exclude the model if any single signal region contains more signals than its 95\% CLs upper limit. Massive gauge bosons are decayed leptonically to improve statistics. We verified that we could reproduce their results with good accuracy. The main background is by far $WZ$ production. Results will be presented in conjunction with those for small $r_{\text{inv}}$ in the next subsection.

\subsection{Case III with small $r_{\text{inv}}$}\label{ssSec:CaseIIILowrinv}
The opposite limit of $r_{\text{inv}}\to 0$ is far less trivial to constrain. We do so by continuing to focus on $N_2 C^-$ pair production and its conjugate process. The neutral fermion $N_2$ is still assumed to decay to a $Z$ and $N_1$. The signature then consists of a $W$, a $Z$ and two jets but no sizable amount of MET. This non-trivial signature does not correspond to any common BSM signature, but can be constrained by measurements of the differential $WZ$ cross section. More specifically, we look at measurements of the $WZ$ cross section that also count the number of jets.

Limits are obtained for small $r_{\text{inv}}$ by recasting the cuts of the measurement of the differential $WZ$ cross sections of Refs.~\cite{Aaboud:2016yus, Khachatryan:2016tgp, Khachatryan:2016poo}. The measurement of Ref.~\cite{Aaboud:2016yus} was performed by ATLAS at $\sqrt{s}=13$~TeV and corresponds to an integrated luminosity of 3.2~fb$^{-1}$. The measurement of Ref.~\cite{Khachatryan:2016tgp} by CMS was also taken at $\sqrt{s}=13$~TeV with an integrated luminosity of 2.3~fb$^{-1}$. Finally, Ref.~\cite{Khachatryan:2016poo} was taken by CMS mainly at $\sqrt{s}=8$~TeV with an integrated luminosity of 19.6~fb$^{-1}$. Ref.~\cite{Khachatryan:2016poo} also includes some results at 7 TeV, but we neglect them because of their minimal integrated luminosity. We verified our results in every case by reproducing with good accuracy their efficiencies for the signal regions associated to the fiducial cross section. A number of 10000 events were generated for each point of the parameter space simulated and for each studied energy. Massive gauge bosons were decayed to leptons at \texttt{MadGraph} level to improve statistics. We obtain by simulation the number of signals with at least two jets for each search. Limits are then obtained by statistically combining together these results using CLs techniques. Correlations between the different searches are neglected as the data was either taken at different energies or different detectors.

\begin{figure}[t]
\begin{center}
    \includegraphics[width=0.6\textwidth, viewport = 0 0 450 400]{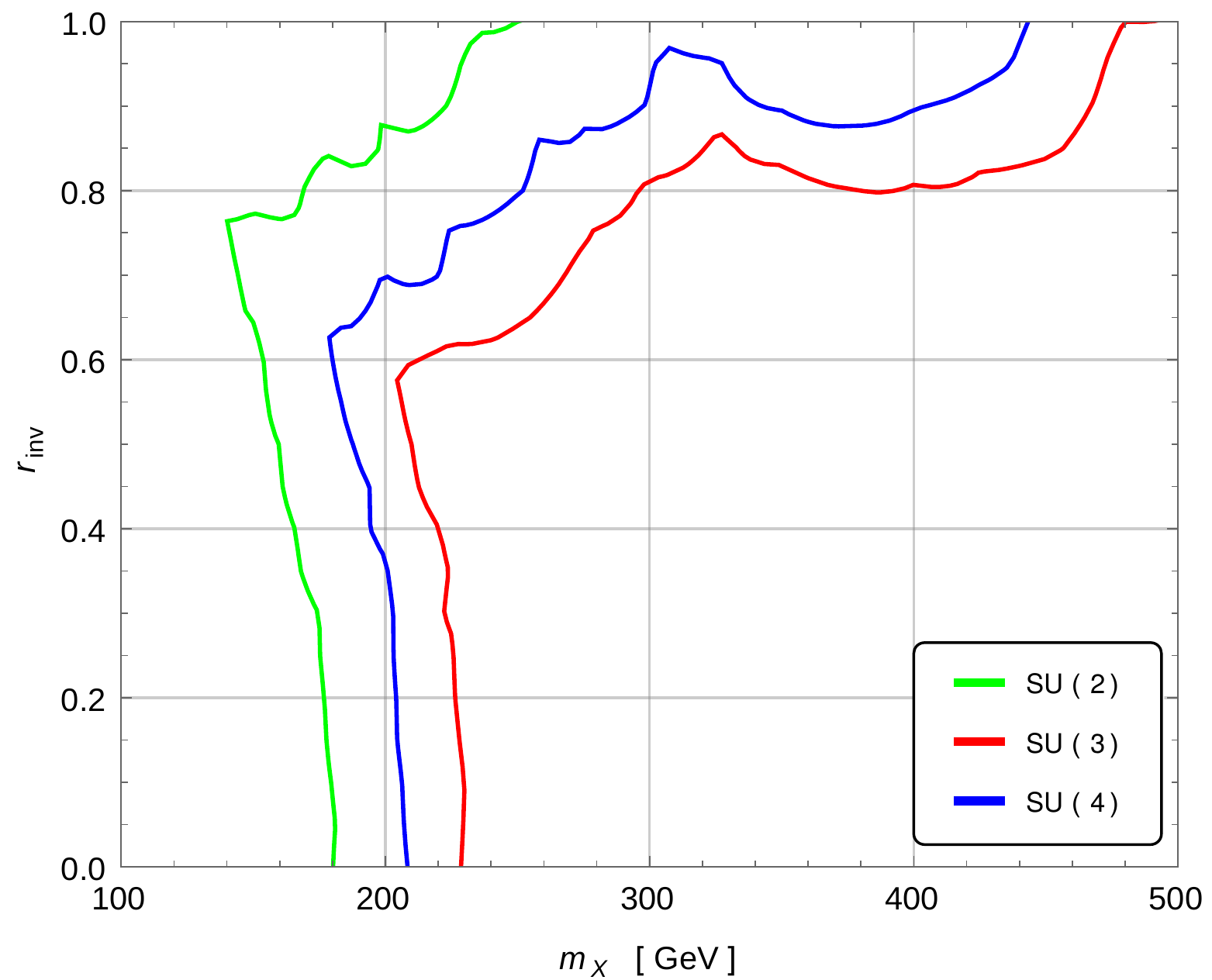}
  \end{center} 
\caption{Summary of the bounds for case III for $\mathcal{G}$ corresponding to $SU(N)$. The bounds at low $r_{\text{inv}}$ come from measurements of the differential $WZ$ cross section and those at large $r_{\text{inv}}$ from Refs.~\cite{Sirunyan:2017qaj} and \cite{Sirunyan:2017lae}. The coefficients were set to $y_{H,L}^F=y_{H,R}^F=0.05$.} \label{fig:CaseIII_summary}
\end{figure}

Constraints on $m_X$ and $r_{\text{inv}}$ are shown in Fig.~\ref{fig:CaseIII_summary} for $\mathcal{G}$ corresponding to different $SU(N)$. This plot includes both the electroweakino bounds of Refs.~\cite{Sirunyan:2017qaj, Sirunyan:2017lae} and those coming from measurements of the differential $WZ$ cross section. A point is excluded only if any of these strategies individually rejects it. As can be seen, the limits are quite good for large $r_{\text{inv}}$ but fall drastically once this parameter drops. There are two reasons why the bounds are smaller than those of Ref.~\cite{Sirunyan:2017qaj} when $r_{\text{inv}}=1$. The first one is that their gauginos are in the adjoint representation of $SU(2)_L$ while the mediators of case II are in the fundamental, which results in a smaller cross section. The second is that our signal is suppressed by a branching ratio of $N_2 \to N_1Z$, which was assumed to be unity in their case. This results in the analysis being barely able to exclude the signal, which also explains the rather rugged aspect of the curves for large $r_{\text{inv}}$.

\subsection{Case III with intermediary $r_{\text{inv}}$}\label{sSec:CaseIIIMediumrinv}
The region of intermediary $r_{\text{inv}}$ is once again the most foreign to current search strategies. A dedicated search would have to look for combinations of leptons or jets that can be reconstructed to $Z$, $W$ or Higgses, additional jets and MET. It is also possible to exploit the nature of semivisible jets to design an additional cut. Like we did in Sec.~\ref{sSec:CaseIIMediumrinv}, we will take advantage of our ability to predict the direction of the MET by determining which jets are most likely to be the semivisible ones.

We once again consider the pair production of $N_2$ and $C^-$ and its conjugate process, with $N_2$ decaying to a $Z$ and $N_1$. We first require the presence of exactly two leptons compatible with the decay of a $Z$ boson. We can then trivially reconstruct the momentum of the $Z$ candidate. We then require the presence of at least four jets. We select the pair that is closest to the $W$ mass in terms of its invariant mass. This grants us the information on the momentum of the $W$ candidate. Finally, we can pair the $Z$ candidate with a different jet and do the same thing with the $W$ candidate. We then keep the pairing that minimizes the difference between the invariant masses. The jets that were selected this way, but which are not associated to the $W$ candidate, are then the most likely semivisible jet candidates. By adding vectorially together their transverse momenta, we obtain the most likely direction for the MET. We can then once again define $\Delta\phi_{\text{MET}}^J$ as the difference between this direction and the direction of the actual MET. The difference between the mass of $C^+$ and $N_2$ is too small to have any noticeable effects.

\begin{figure}[t]
\begin{center}
    \includegraphics[width=0.6\textwidth, viewport = 0 0 450 400]{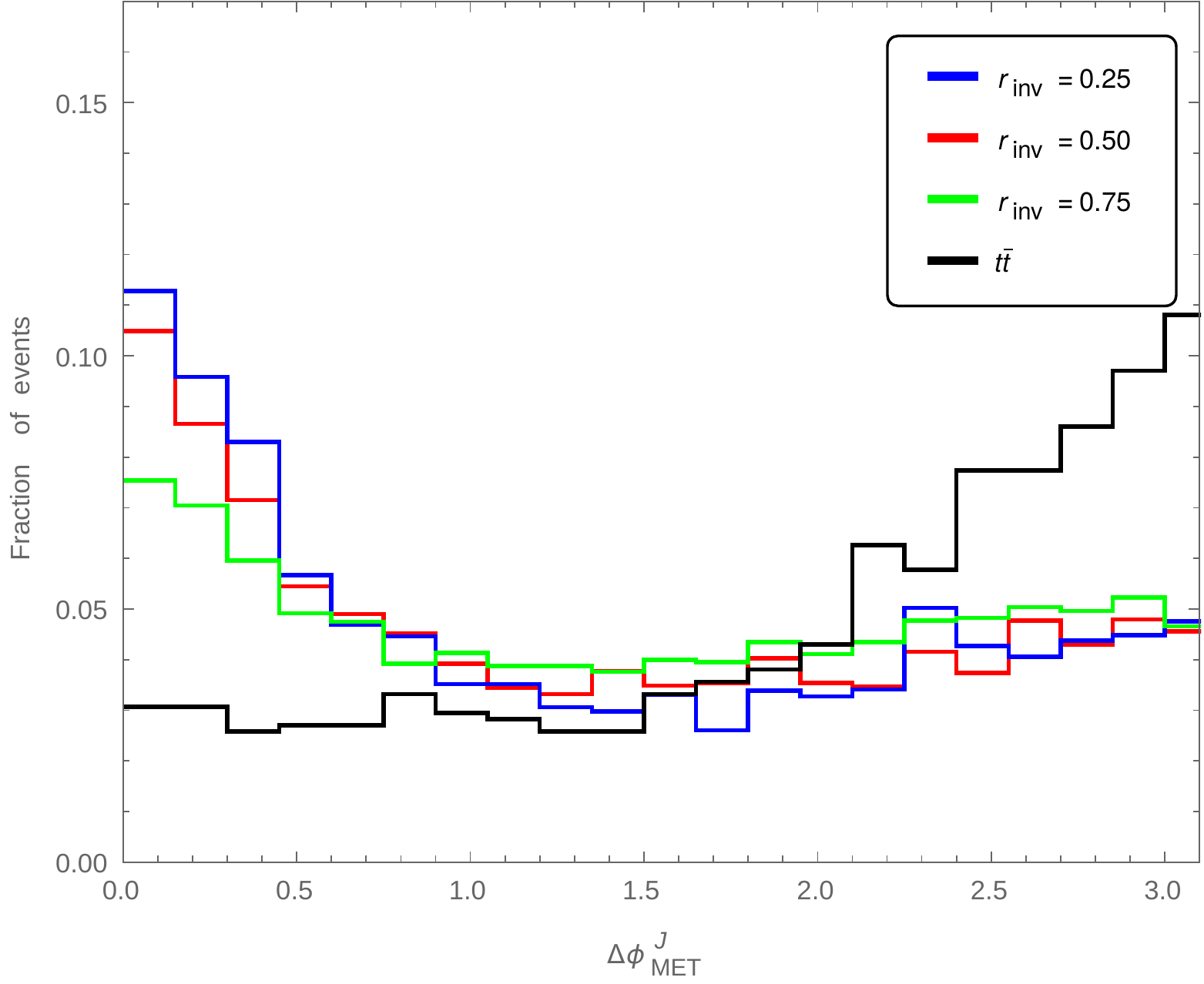}
  \end{center} 
\caption{Normalized distribution of $\Delta\phi_{\text{MET}}^J$ in case III for different signals and the $t\bar{t}$ background. The mass $m_X$ was taken to be 500~GeV and the coefficients were set to $y_{H,L}^F=y_{H,R}^F=0.05$. Cuts were applied beforehand as described in the text.} \label{fig:New_cut_2}
\end{figure}

The distribution of $\Delta\phi_{\text{MET}}^J$ is shown in Fig.~\ref{fig:New_cut_2} for sample signals and the $t\bar{t}$ background. 
The cuts of the electroweakinos search of Ref.~\cite{Sirunyan:2017qaj} were applied but with a few modifications. First, the requirement of a minimal separation between the jets and the MET was removed for obvious reasons. Second, the upper limit on the invariant mass of the two closest jets was replaced by the requirement of two jets having an invariant mass between 75 and 85~GeV. Third, we replaced the different MET signal regions by a simple requirement of at least 100~GeV. Finally, the presence of at least four jets was required. A number of $10^5$ signals and 5 millions $t\bar{t}$ background were generated. The $Z$ and $W$ were respectively decayed to leptons and hadrons at \texttt{MadGraph} level to improve statistics. The veto on b-tagged jets and the cut on the stransverse mass were also turned off for the sake of improving the quality of the histogram, though of course a real search would never do this. Neither of these two cuts affect the plot qualitatively. As can be seen, the distribution of $\Delta\phi_{\text{MET}}^J$ peaks at 0, while $t\bar{t}$ peaks around $\pi$. The peak for the signal is not as clear as in Sec.~\ref{sSec:CaseIIMediumrinv}. This is simply a consequence of the considerably more complicated procedure for identifying the candidate semivisible jets and the manipulation of more jets, which are inherently less clean objects than leptons. We leave the details of a complete optimization of a search strategy and its prospects for future work.

\section{Case IV: Operators linear in $X$ involving only scalars}\label{Sec:CaseIV}
Operators of category IV are those that are linear in $X$ and only contain scalars. The four possibilities are:
\begin{equation}\label{eq:OperatorCaseIV}
  \begin{aligned}
    &\lambda_{4s1}  H^\dagger (n^S)^\dagger X^S_{4s1} + \text{h.c.} & \;\;\;\;\;
    &\lambda_{4s2} H^\dagger (n^S)^\dagger X^S_{4s2} n^S +\text{h.c.}\\
    &\lambda_{4s3} (n^S)^\dagger H^\dagger X^S_{4s3} H +\text{h.c.} &
    &\lambda_{4s4} (n^S)^\dagger H^T X^S_{4s4} H +\text{h.c.}
  \end{aligned}
\end{equation}
In many ways, their phenomenology is similar to that of a conventional extended Higgs sector. More precisely, the mediators resemble fermiophobic Higgses. As explained in Sec.~\ref{sSec:GeneralAssumptions}, we assume that $\mathcal{G}$ is unbroken in the absence of these operators. Sufficiently large coefficients will however lead to $X$ and $n$ acquiring non-zero expectation values and thus to $\mathcal{G}$ being broken. In this section, we will consider both the possibilities of $\mathcal{G}$ being broken or not. To facilitate the discussion, we will assume $\mathcal{G}$ to be $SU(N)$, but the results are essentially independent of this choice. We refer to appendix~\ref{app:details} for details on the embedding of the representations of the unbroken subgroup inside the representations of $\mathcal{G}$. The main conclusion will be that current search strategies are not very sensitive to these mediators, but that some of these particles could be better constrained by looking at some understudied signatures.

\subsection{Operator $\lambda H^\dagger n^\dagger X + \text{h.c.}$}\label{sSec:CaseIVa}
The mediator $X^S_{4s1}$ consists of a doublet of $SU(2)_L$ of weak hypercharge $+1/2$. Its simplest representation under $\mathcal{G}$ is that of a fundamental, which we assume from now on. In this case, $n^S$ is also forced to be a fundamental of $\mathcal{G}$.

When neither the mediator nor $n^S$ acquire an expectation value, the collider phenomenology is trivial. The Higgs acquiring an expectation value leads to the neutral part of $X^S_{4s1}$ mixing with $n^S$. The resulting scalar spectrum is then a charged scalar $C^+$ and two neutral scalars $N_1$ and $N_2$, with the latter being the heaviest. The charged scalar decays mainly via $C^+ \to N_1 W^+$. The scalar $N_1$ is stable while the other neutral scalar decays either via $N_2 \to N_1 Z$ or $N_2 \to N_1 h$. These scalars are of course always pair produced. At this point, it is clear that the phenomenology is the same as that of case III. The only difference is that the cross section is much smaller. We verified that our bounds vanish for such small cross sections and for groups of reasonable rank.

The alternative is that $X^S_{4s1}$ and $n^S$ acquire expectation values, thereby breaking $\mathcal{G}$ to an unbroken subgroup $\mathcal{H}$. Assuming $\mathcal{H}$ to be non-trivial, the resulting dark gauge bosons fall into three categories. First, there are massless gauge bosons $g_D^U$ in the adjoint representation of $\mathcal{H}$. Second, some of them form a fundamental representation of massive complex gauge bosons $\hat{g}_D^U$. Finally, there is a singlet massive boson $g_D^B$. Obviously, the latter mixes with $W^3$ and the weak hypercharge boson to create a new massive gauge boson $U$. This has sizable effects on EWPT, much more so than if $X^S_{4s1}$ had been a singlet under $\mathcal{G}$. It would be possible to adjust the vev of $n^S$ to minimize mixing between $g_D^B$ and electroweak gauge bosons, but the vev of $X^S_{4s1}$ could at best be of $\mathcal{O}(10)$~GeV without sizable tuning of the potential. Of course, the exact limit would require a more elaborate study of EWPT in conjunction with a study of the potential and is far beyond the scope of this article. Note also that the phenomenology of $U$ is similar to that of U-bosons \cite{Fayet:1980ad}. These are already considerably constrained (see for example \cite{Alves:2016cqf, Pospelov:2008zw, Aranda:1998fr, Fayet:2007ua, Curciarello:2015iea, Curciarello:2015xja}), though we verified that the coupling of $U$ to Standard Model fermions are typically below their limits. Of course, measurements of the Higgs signal strengths and branching ratios would also put considerable constraints on the vev of the mediator. The main point however is that the vev of $X^S_{4s1}$ is forced to be small and this suppresses the cross section for the production of a single mediator to values far below the typical experimental constraints.

The spectrum of physical scalars consists of particles that are either fundamentals or singlets under $\mathcal{H}$. There are charged and neutral fundamentals of $\mathcal{H}$ labeled respectively $X'^U_+$ and $X'^U_0$. There are also the equivalent singlets $X'^B_+$ and $X'^B_0$. Finally, there is a CP-even real singlet $n'^B$, which is typically light. Of course, some scalars have been eaten by the massive dark gluons and some of the mass eigenstates are linear combinations of components of either $X^S_{4s1}$, $n^S$ and $H$. The scalars charged under $\mathcal{H}$ decay mainly via $X'^U_+ \to \hat{g}^U_D W^+$ and $X'^U_0 \to \hat{g}^U_D x$, where $x$ is either $Z$, $U$, $h$, $n'^B$ or possibly $X'^B_0$ if kinematically allowed. The scalars that are singlet under $\mathcal{H}$ however have a vast variety of possible decay channels as they mix with the components of the Standard Model Higgs doublet. Unfortunately, pair production of scalars leads to either signatures that are equivalent to those for an unbroken $\mathcal{G}$ or new signatures that would be even more difficult to discover. This is rather unsurprising considering the Goldstone boson equivalence theorem \cite{Cornwall:1974km, Vayonakis:1976vz} and the constraints on the vev of $X^S_{4s1}$. All in all, there are practically no collider limits on $X^S_{4s1}$ even when $\mathcal{G}$ is broken.

\subsection{Operator $\lambda H^\dagger n^\dagger X n + \text{h.c.}$}\label{sSec:CaseIVb}
The mediator $X^S_{4s2}$ is again a doublet of $SU(2)_L$ of weak hypercharge $+1/2$. The main difference with respect to $X^S_{4s1}$ is its gauge numbers under $\mathcal{G}$. The simplest possibility and the one which we will assume from now on is that $X^S_{4s2}$ is an adjoint of $\mathcal{G}$. The dark squark $n^S$ is taken to be a fundamental.

When $\mathcal{G}$ is unbroken, the collider phenomenology is again very straightforward. Because they transform under different representations of $\mathcal{G}$, there is no mixing between the components of $X^S_{4s2}$ and $n^S$. The neutral component of $X^S_{4s2}$, which we label $X^0$, decays via $X^0 \to \bar{n} n$. The charged component is labeled $X^+$ and decays via the three-body process $X^+ \to \bar{n} n W^+ $. The two-body decay $X^+ \to g_D W^+$ is not forbidden by gauge symmetry, but is not generated at either tree-level or at one loop and is therefore negligible. By far the easiest signature to detect would be the pair production of two $X^+$. This would lead to a pair of $W$ bosons and four semivisible jets. Unfortunately, there are no recent experimental searches that cover this signature efficiently. The other possibilities would simply not be spectacular enough to distinguish them from the background. 

When $\mathcal{G}$ is broken to $\mathcal{H}$, the dark gluons sector remains the same as in Sec.~\ref{sSec:CaseIVa}. The scalar sector is however extended. There is both a charged and neutral adjoint of $\mathcal{H}$ labeled respectively $X'^U_+$ and $X'^U_0$. There are two fundamentals of $\mathcal{H}$ each containing a charged and neutral field labeled $\hat{X}'^U_{i+}$ and $\hat{X}'^U_{i0}$ respectively, with $i$ being either 1 or 2. There are also new charged and neutral singlets under $\mathcal{H}$ referred to as $X'^B_+$ and $X'^B_0$ respectively. Finally, there is a CP-even singlet $n'^B$. The leading decays for the scalars charged under $\mathcal{H}$ are:
\begin{equation}\label{eq:CaseIVbdecay}
  \begin{aligned}
    & X'^U_+ \to W^+\;\hat{g}_D^U\; \bar{\hat{g}}_D^U & \;\;\;\;\; & X'^U_0 \to \hat{g}_D^U\; \bar{\hat{g}}_D^U \\
    & \hat{X}'^U_{i+} \to W^+\;\hat{g}_D^U\;U/n'^B                  &     & \hat{X}'^U_{i0} \to \hat{g}_D^U\; U/n'^B.
  \end{aligned}
\end{equation}
Other decay channels would become important if the vev of $X^S_{4s2}$ were allowed to be large. The decays for the scalars not charged under $\mathcal{H}$ are more complicated because of mixing with the Higgs doublet. The collider signatures are similar to those for unbroken $\mathcal{G}$ and the mediators are not expected to be any more constrained.

\subsection{Operator $\lambda n^\dagger H^\dagger  X H + \text{h.c.}$}\label{sSec:CaseIVc}
The mediator $X^S_{4s3}$ consists of a triplet of $SU(2)_L$ with no weak hypercharge. Its simplest representation under $\mathcal{G}$ is that of a fundamental, which we will assume from now on. In this case, $n^S$ is also forced to be a fundamental of $\mathcal{G}$.

When $\mathcal{G}$ is unbroken, the collider phenomenology is not so different from that of Sec.~\ref{sSec:CaseIVa}. The main difference is that there are now positively charged, neutral and negatively charged fundamentals of $\mathcal{G}$. A possible signature to study would then be production of two same sign mediators via $W$ vector boson fusion. The signature would then be two same sign $W$, two jets and two semivisible jets. This is specially interesting for the possibility of two same sign leptons in conjunction with MET and high jet multiplicity.  We are unaware of any recent experimental search that would be optimized for such a signal with comparable cross sections.

When $\mathcal{G}$ is instead broken, the vev of the mediator is even more strongly constrained than for $X^S_{4s1}$ and $X^S_{4s2}$. The reason is that, even if $X^S_{4s3}$ were not charged under $\mathcal{H}$, it would still break custodial symmetry and as such its vev could only be at best a few GeV. Do note that the broken dark gluon does not mix at tree level with $W^3$ and the weak hypercharge boson. Considering the constraint on the vev, the signatures should be equivalent to those for an unbroken $\mathcal{G}$.

\subsection{Operator $\lambda n^\dagger H^T  X H + \text{h.c.}$}\label{sSec:CaseIVd}
The mediator $X^S_{4s4}$ is also a triplet of $SU(2)_L$ but with a weak hypercharge of $-1$ instead. Its simplest representation under $\mathcal{G}$ is again a fundamental, in which case $n^S$ must also be one.

When $\mathcal{G}$ is unbroken, the scalar content consists of fundamentals of $\mathcal{G}$ that are either neutral, charged or doubly-charged. The most spectacular signature would without doubt be the pair production of two doubly charged mediators, which would lead to a total of four $W$ and two semivisible jets. To the best of our knowledge, there are no current search which would be optimized for such a signature.

Since $X^S_{4s4}$ would break custodial symmetry even if it were not charged under $\mathcal{G}$, its vev is constrained to be small when $\mathcal{G}$ is broken. As such, the signatures are in good approximation equivalent to those for an unbroken $\mathcal{G}$.

\section{Case V: Operators non-linear in $X$}\label{Sec:CaseV}
Operators of category V are those that are non-linear in $X$ and do not preserve any discrete symmetry that would maintain the mediators stable. They are:
\begin{equation}\label{eq:OperatorCaseV}
  \begin{gathered}
    \lambda_{nl1}X^S_{nl1}X^S_{nl1}(X^S_{nl1})^\dagger H + \text{h.c.}\\
    \lambda_{nl2}X^S_{nl2}X^S_{nl2}(X^S_{nl2})^\dagger n + \text{h.c.}\\
    \lambda_{nl3}X^S_{nl3}X^S_{nl3}(X^S_{nl3})^\dagger   + \text{h.c.}
  \end{gathered}
\end{equation}
These operators are a bit of an oddity. First, it is hard to think of examples of theories in which they could appear. In addition to their rather strange structure, they require complicated combinations of gauge numbers to be at the same time both gauge invariant and non-zero. It is a real possibility that these operators are simply a theoretical curiosity and nothing more. Second, they allow for a multitude of possible combinations of gauge numbers, including the possibility of carrying color or not. This renders a complete analysis of their constraints or prospects at best a daunting task. Third, they sometimes lead to very unconventional signatures that cannot be efficiently constrained using current experimental results.

Because of this, we limit ourselves to presenting the simplest case for each operator and describing their experimental signatures.

\subsection{Operator $\lambda XXX^\dagger H + \text{h.c.}$}\label{sSec:CaseVa}
The simplest possibility for the gauge numbers of $X^S_{nl1}$ is that it be a doublet of $SU(2)_L$ of weak hypercharge $-1/2$ and an adjoint of $\mathcal{G}$. This is valid as long as $\mathcal{G}$ is compact. For the sake of simplicity, we temporarily drop the subscript and superscript of $X$. We replace them by an $SU(2)_L$ index ranging from 1 to 2 and a $\mathcal{G}$ index ranging from 1 to the size of the adjoint representation of $\mathcal{G}$. We label the $SU(2)_L$ indices by Greek letters and the $\mathcal{G}$ indices by Roman ones. We can then define
\begin{equation}\label{eq:CaseVaX}
  X^\alpha=X^{a\alpha}t_a,
\end{equation}
where the $t_a$ are the generators of $\mathcal{G}$ normalized such that $\text{Tr}[t_a t_b]=\delta_{ab}/2$. The operator can therefore be written as
\begin{equation}\label{eq:CaseVaOp1}
  \mathcal{L} = \lambda \epsilon_{\alpha\beta}\text{Tr}\left[X^\alpha X^\beta X^\dagger_\gamma\right]H^\gamma+\text{h.c.},
\end{equation}
where we dropped the indices on $\lambda$ as well. We can then expand the $SU(2)_L$ indices and express the operator in terms of the neutral and charged components of the mediator $X_0$ and $X_-$ respectively. This gives:
\begin{equation}\label{eq:CaseVaOp2}
  \mathcal{L} = \frac{i}{2}\lambda f_{ab}{}^c \left(X_0^a X_-^b X_{0c}^\dagger H^+ +  X_0^a X_-^b X_{-c}^\dagger H^0 \right) + \text{h.c.},
\end{equation}
where $f_{ab}{}^c$ are the structure constants of $\mathcal{G}$. A very large value of $\lambda$ would lead to the breaking of electromagnetism, though the exact value at which this happens depends on the exact details of $\mathcal{G}$. Below this value, $X$ does not acquire a vev and $\mathcal{G}$ remains unbroken. In this case, the first term of (\ref{eq:CaseVaOp2}) can be removed by passing to the unitary gauge.

\begin{figure}[t!]
  \centering
  \begin{subfigure}{0.42\textwidth}
    \centering
    \includegraphics[width=\textwidth, viewport = 0 0 260 180]{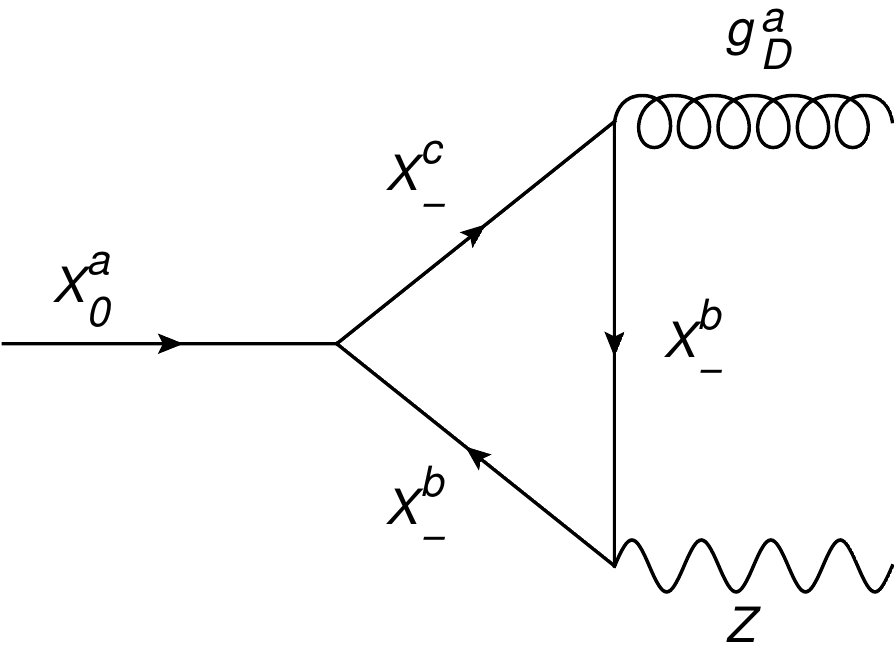}
    \caption{}
    \label{fig:Feyn1}
  \end{subfigure}
  ~
    \begin{subfigure}{0.42\textwidth}
    \centering
    \includegraphics[width=\textwidth, viewport = 0 0 260 180]{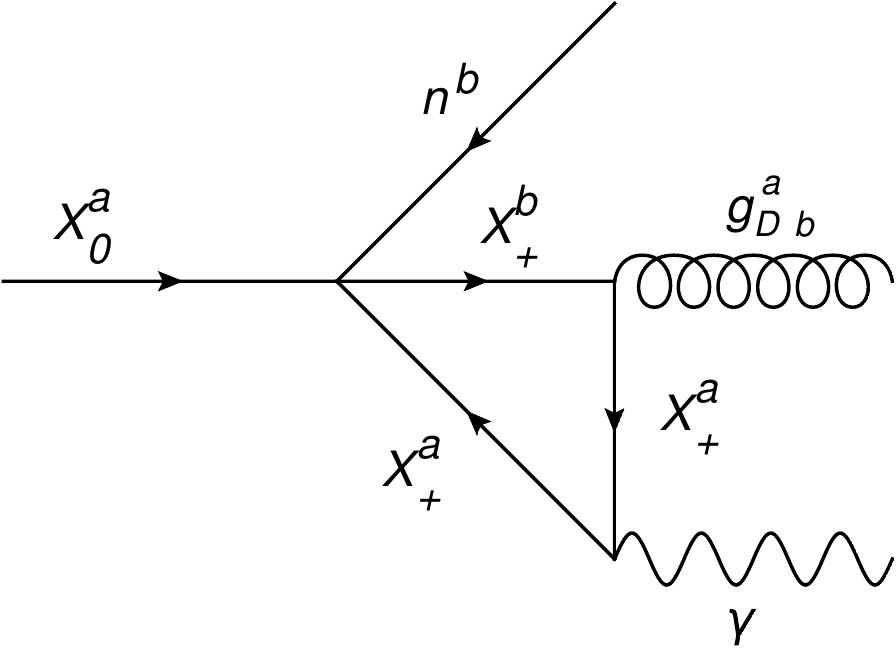}
    \caption{}
    \label{fig:Feyn2}
  \end{subfigure}
\caption{(a) Example decay of a neutral $X^S_{nl1}$. (b) Example decay of a neutral $X^S_{nl2}$. The dark color index is in the adjoint representation for the first case and in the fundamental for the second.}\label{fig:CaseV}
\end{figure}

At colliders, $X^S_{nl1}$ would mainly be pair produced via an $s$-channel photon, $Z$ or $W$. The mediator will subsequently decay to a pair of gauge bosons, which can be any dark gluons, photons, $Z$ or $W$. The only restrictions are that the decay products always contain at least one dark gluon and that electric charge be conserved. An example is shown in Fig.~\ref{fig:Feyn1}. The experimental signature is MET, $2+n$ semivisible jets and $2-n$ gauge bosons $W/Z/\gamma$, with $n$ being any integer value between 0 and 2 inclusively. Do note that the complicated nature of this operator would lead to the decay width of the lightest dark hadron being massively suppressed. In light of other constraints, it might not be reasonable to require the dark hadrons to decay promptly. We leave the details of this question for future work.

\subsection{Operator $\lambda XXX^\dagger n + \text{h.c.}$}\label{sSec:CaseVb}
The simplest possibility for the gauge numbers of $X^S_{nl2}$ is that it be a fundamental of $\mathcal{G}$ and an adjoint of $SU(2)_L$ with no weak hypercharge. The dark squark $n^S$ can then be an antifundamental of $\mathcal{G}$. Using the same notation as in the last subsection, one can define
\begin{equation}\label{eq:CaseVbX}
  X^a=X^{a\alpha}\sigma_\alpha/2.
\end{equation}
The operator can then be written as
\begin{equation}\label{eq:CaseVbOp1}
  \mathcal{L} = \lambda \text{Tr}\left[X^a X^b X^\dagger_a\right]n_b+\text{h.c.},
\end{equation}
where we also dropped the superscript on $n$. For each dark color, $X$ consists of three particles: a positively charge $X_+$, a neutral $X_0$ and a negatively charged $X_-$. Expanding the $SU(2)_L$ indices leads to
\begin{equation}\label{eq:CaseVbOp2}
  \begin{aligned}
    \mathcal{L} = & \frac{\lambda}{4}\left(X_+^a X_-^b - X_-^a X_+^b\right)X^\dagger_{0a} n_b +\text{h.c.}\\
                + & \frac{\lambda}{4}\left(X_-^a X_{-a}^\dagger - X_+^a X_{+a}^\dagger  \right)X^b_0 n_b +\text{h.c.}\\
                + & \frac{\lambda}{4}\left(X_+^b X_{+a}^\dagger - X_-^b X_{-a}^\dagger  \right)X^a_0 n_b +\text{h.c.}
  \end{aligned}
\end{equation}
Like in Sec.~\ref{sSec:CaseVa}, this operator does not lead to $X$ acquiring an expectation value unless it also breaks the electromagnetic group. The group $\mathcal{G}$ must therefore remain unbroken.

At colliders, the mediator $X^S_{nl2}$ will once again mainly be pair produced via an $s$-channel photon, $Z$ or $W$. Its decay is however less trivial. The fact that $X$ is a fundamental of $\mathcal{G}$ means that its decay products must include at least one $n^\dagger$. The decay can either be three-body at one loop or two-body at two loops. The three-body decay is to $n^\dagger$, a dark gluon and either a $Z$, $W$ or photon depending on the charge of the decaying mediator. Decays involving two SM gauge bosons or two dark gluons are not generated at one loop. The two-body decay is to $n^\dagger$ and either a $Z$, $W$ or photon. Two-body decays to dark gluons or two SM bosons are not generated at one loop. An example is shown in Fig.~\ref{fig:Feyn2}. The signature for pair production then consists of two SM gauge bosons and 2 to 4 semivisible jets.

\subsection{Operator $\lambda XXX^\dagger + \text{h.c.}$}\label{sSec:CaseVc}
The simplest combination of gauge numbers that $X^S_{nl3}$ can take is it being an adjoint of both $\mathcal{G}$ and $SU(2)_L$ with 0 weak hypercharge. We can define
\begin{equation}\label{eq:CaseVcX}
  X=X^{a\alpha}t_a\sigma_\alpha/2.
\end{equation}
The operator is then written as
\begin{equation}\label{eq:CaseVcOp1}
  \mathcal{L} = \lambda \text{Tr}\left[X X X^\dagger\right]+\text{h.c.},
\end{equation}
where the trace is taken over both $SU(2)_L$ and $\mathcal{G}$ indices. Once expanded over $SU(2)_L$ indices, it gives
\begin{equation}\label{eq:CaseVcOp2}
  \mathcal{L} = i\frac{\lambda}{8}f_{ab}{}^c\left(X^a_+X^b_-X^\dagger_{0c} + X^a_0X^b_+X^\dagger_{+c} + X^a_- X^b_0X^\dagger_{-c} \right)+\text{h.c.}
\end{equation}
Like with the two others, this operator does not lead to $X$ acquiring an expectation value unless it also breaks electromagnetism. The mediator decays mainly to a dark gluon and either a $W$, $Z$ or photon depending on its charge. Diagrams are similar to Fig.~\ref{fig:Feyn2} with some small differences. The signature for pair production of mediators is then 2 SM gauge bosons and 2 semivisible jets.

\section{Conclusion}\label{Sec:Conclusion}
As the LHC continues to accumulate new data, the most common BSM theories will have to hide in an increasingly small region of parameter space if they are to remain potentially discoverable. This motivates the search for more exotics signals of which Hidden Valley models are a great source. These models are characterized by a new confining sector that communicates with the Standard Model via heavy mediators. Up to now, the vast majority of the work on such particles has concentrated on mediators of spin 1, with other possibilities being vastly neglected.

In this work, we investigated direct production of Hidden Valley mediators of spin 0 or 1/2 that are non-trivially charged under both the Standard Model gauge groups and a single new confining group. Because of their exotic signatures, we mainly focused on semivisible jets, objects which look like normal jets but which are accompanied by collinear MET. More precisely, we sought to provide an overview of the current collider constraint on these mediators and show that vast regions of parameter space are still relatively unconstrained. We also developed new techniques to better constrain semivisible jets by exploiting the somewhat unusual positioning of the MET with respect to other objects.

For the mediators to be able to decay, a global $U(1)$ symmetry must be broken. This is best done by the introduction of operators that explicitly break this symmetry. We have shown that there is only a finite number of such renormalizable operators and that they can be classified into five distinct categories. The first one consists of three field operators involving a quark. Depending on the fraction $r_{\text{inv}}$ of the energy of the semivisible jet that is transmitted to MET, this scenario can be well constrained by searches for squark pair production or searches for paired dijets. Intermediary values of $r_{\text{inv}}$ are relatively less constrained given current search strategies, but could be better probed by dedicated searches. Case II consists of three field operators involving leptons. Large values of $r_{\text{inv}}$ are well constrained by slepton searches, but small values are still relatively unconstrained due to the fact that an electroweak signal competes with many strong backgrounds. Dedicated searches could however greatly increase the constraints on intermediary values of $r_{\text{inv}}$ by exploiting the ability to reliably determine which jets in an event are the most likely semivisible jet candidates. Operators of category III are three field operators involving the Higgs doublet, a fermion mediator and a dark quark. At large $r_{\text{inv}}$, they are relatively well constrained by electroweakino searches, while regions of small $r_{\text{inv}}$ are only weakly constrained by measurements of the differential $WZ$ cross section. Intermediary values of $r_{\text{inv}}$ could be further probed by dedicated searches by again using our ability to infer which jets are the most likely semivisible jet candidates. Case IV consists of operators linear in the mediator and involving only scalars. They generally lead to breaking of the new confining group and are mostly constrained by Higgs physics and EWPT. Case V consists of the operators that are non-linear in the mediator. Although they might simply be a theoretical curiosity, they lead to very exotic signatures some of which are virtually unsearched for.

Although we believe we provided a vast amount of information on scalar and fermion mediators, the present article mainly focused on collider signatures while neglecting many other constraints. Due to the ubiquity of such mediators in the literature, dedicated studies on other aspect of these mediators would certainly be justified. Namely, constraints from electroweak precision tests and flavor physics would need to be more properly studied. Perhaps more importantly, a more systematic study of how these mediators could fit in the context of non-abelian dark matter would be of significant importance.

\acknowledgments
This work was supported by Fundação de Amparo à Pesquisa (FAPESP), under contracts 16/17040-3, 16/17041-0, 16/02636-8 and 15/25884-4, and Conselho Nacional de Ci\^{e}ncia e Tecnologia (CNPq). We are particularly grateful to Oscar J. P. \'Eboli for discussions and collaboration at the early stage of the project. We would like to thank the CMS collaboration for making public some previously unpublished numbers from Ref.~\cite{Khachatryan:2016poo}. We would also like to thank the participants of the workshop {\it LHC Chapter II: The Run for New Physics} for useful discussions. GGdC is grateful to the University of Southampton, the Laboratoire de Physique des Hautes \'Energies, GRAPPA, the University of Mainz and the DESY theory group for kind hospitality at various stages of this work.

\appendix
\section{Some details on the dark gauge bosons and scalar fields}\label{app:details}
We will now present some details on the dark gauge bosons and scalar fields of Sec.~\ref{Sec:CaseIV}. For ease of presentation, we will drop all the subscripts and superscripts, denoting the mediator simply by $X$ and the light scalar field by $n$. When both $X$ and $n$ acquire a vev, the dark gauge group $\mathcal{G}$ will be spontaneously broken to a subgroup $\mathcal{H}$. From now on we will take $\mathcal{G} = SU(N)$ and $\mathcal{H} = SU(N-1)$. The matrix of the $\mathcal{G}$ gauge bosons can be decomposed as
\begin{align}\label{eq:dark_gbosons}
\begin{aligned}
  g_D  = \left(
  \begin{array}{ccc|c}
    & & & \\
    & g_D^U + \sqrt{\frac{2}{N(N-1)}}  g_D^B &  & \hat{g}_D^U \\ 
    & & & \\\hline\\[-1.0em]
    & \hat{g}_D^{U\dag} & & - \sqrt{\frac{2(N-1)}{N}} g_D^B
  \end{array}
  \right).
\end{aligned}
\end{align}
The $N(N-2)$ massless vectors of the unbroken subgroup $\mathcal{H}$ are denoted by $g_D^U$. In addition, there are $2(N-1)$ massive vectors $\hat{g}_D^U$ transforming in the fundamental of $\mathcal{H}$ and a massive singlet $g_D^B$. Notice that an identity $\mathbbm{1}_{N-1}$ accompanying the singlet vector is left implicit in the upper left block. Depending on the $SU(2)_L \times U(1)_Y$ gauge numbers of the mediator, $g_D^B$ may mix with the $Z$ boson via the vev $v_X$, generating the physical vector boson $U$. Notice that the mixing between $g_D^B$ and $Z$ constrains the vev of the mediator to be below a few GeV to guarantee compatibility with data. 

Turning to the mediator, we have seen in Sec.~\ref{Sec:CaseIV} that, in the simplest cases, $X$ will be either a fundamental or an adjoint of the dark group $\mathcal{G}$ (although other representations are possible). In the case of a fundamental, we write
\begin{align}\label{eq:mediator_repr1}
  X^{\mathcal{G}-Fund} &\approx \left(
  \begin{array}{c}
  \\
  X'^U \\ 
  \\ \hline\\[-1.0em]
  X'^B
  \end{array} \right),
\end{align}
where $X'^U$ is a fundamental of $\mathcal{H}$ and $X'^B$ a singlet and we use the $\approx$ symbol to significate that there is some admixture with other fields of equivalent gauge numbers. When $X$ is instead an adjoint of $\mathcal{G}$, we write
\begin{align}\label{eq:mediator_repr2}
  X^{\mathcal{G}-Adj} &\approx \left(
  \begin{array}{ccc|c}
  & & & \\
  & X'^U + \sqrt{\frac{2}{N(N-1)}} X'^B &  & \hat{X}'^U_{1} \\ 
  & & & \\\hline\\[-1.0em]
  & (\hat{X}'^U_{2})^\dagger & & - \sqrt{\frac{2(N-1)}{N}} X'^B
  \end{array}
  \right).
\end{align} 
In this case, $X^{\mathcal{G}-Adj}$ can be decomposed in an adjoint of $\mathcal{H}$ ($X'^U$), two fundamentals of $\mathcal{H}$ ($\hat{X}'^U_{1}$ and $\hat{X}'^U_{2}$) and again a singlet ($X'^B$). Of course, in both cases it is only the electrically neutral component of the singlet that acquires a vev $v_X$ and can mix with the Higgs and $n$ scalars. We remind the reader that, although we never wrote it explicitly, each scalar field has an associated $SU(2)_L$ index.

Combining the explicit expressions given in Eqs.~(\ref{eq:dark_gbosons}) and (\ref{eq:mediator_repr1})-(\ref{eq:mediator_repr2}), it is simple to verify the possible mediator decays listed in Sec.~\ref{Sec:CaseIV}.

\section{Effects of the variation of the pion mass}\label{app:variations}
\begin{figure}[t]
  \begin{center}
    \includegraphics[width=0.6\textwidth, viewport = 0 0 450 400]{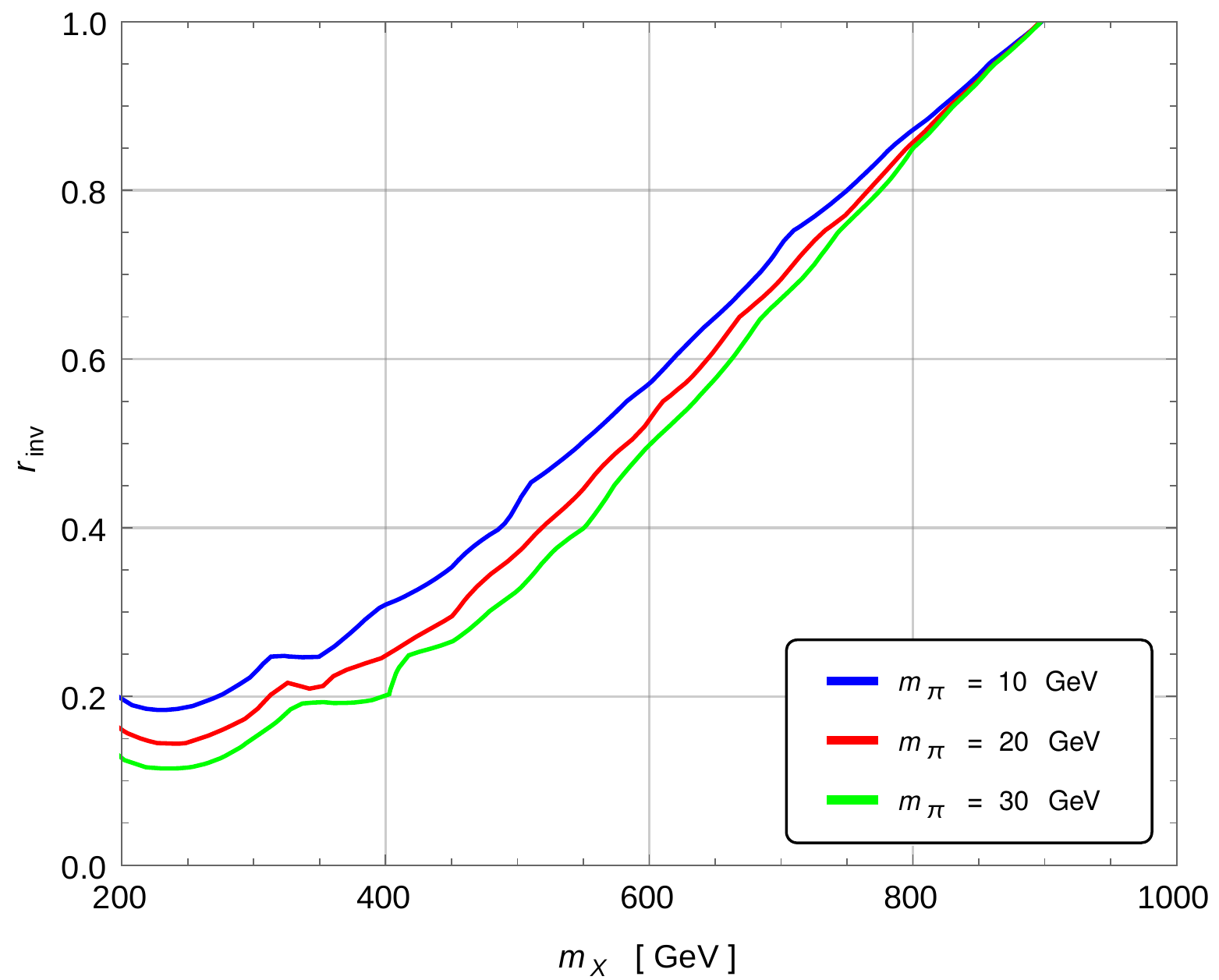}
  \end{center} 
\caption{Bounds at 95\% CLs on $X_L^F$ for different dark pion masses for large $r_{\text{inv}}$. Bounds are obtained by recasting the slepton search of Ref.~\cite{ATLAS:2017uun} and assuming the mediator communicates only with the first generation. The Hidden Valley group $\mathcal{G}$ is taken to be $SU(3)$. The excluded regions are left of the curves.} \label{fig:AnnexII}
\end{figure}

To illustrate the effect of varying the mass of the dark pions, we reran the analysis of Sec.~\ref{sSec:CaseIISmallrinv} for different dark pion masses and $X_L^F$. The results are shown in Fig.~\ref{fig:AnnexII}. The confinement scale was set equal to the dark pion mass. Variations on the bounds are typically smaller for heavier mediators.

\bibliography{biblio}
\bibliographystyle{JHEP}

\end{document}